\pdfoutput=1
\documentclass[iop,apj]{emulateapj}
\usepackage{xspace} 
\usepackage{graphicx,epsfig}
\usepackage{rotate}
\usepackage{amsmath, amsthm, amssymb}
\usepackage{comment} 
\usepackage{times}
\usepackage{soul}
\usepackage{natbib}
\usepackage[plainpages=false, colorlinks=true, anchorcolor=blue,
linkcolor=blue, citecolor=blue, bookmarks=false]
{hyperref}

\DeclareMathAlphabet{\mathitbf}{OML}{cmm}{b}{it}

\newcommand{\code}[1]{{\tt#1}}

\newcommand{\ie}{\emph{i.e.}}   
\newcommand{\eg}{\emph{e.g.}}   

\newcommand{\bq}{\begin{equation}}
\newcommand{\eq}{\end{equation}}

\newcommand{\Erad}{E_\mathrm{r}}

\newcommand{\Etot}{E_\mathrm{tot}}
\newcommand{\Pradeff}{P_\mathrm{rad\,eff}}
\newcommand{\Prad}{P_\mathrm{rad}}
\newcommand{\Emat}{E_\mathrm{m}}
\newcommand{\emat}{e_\mathrm{m}}
\newcommand{\pmat}{p}  
\newcommand{\ptot}{P_\mathrm{tot}}      
\newcommand{\pcorr}{p_\Lambda}      
\newcommand{\vv}{{\mathitbf{v}}}
\newcommand{\kappaP}{\kappa_{\rm P}}
\newcommand{\kappaR}{\kappa_{\rm R}}

\def\gtsim{\lower.5ex\hbox{$\buildrel > \over\sim$}}
\def\ltsim{\lower.5ex\hbox{$\buildrel < \over\sim$}}

\def\sun{\mbox{$_\odot$}}
\def\apjl{ApJL}

\def\apjs{ApJS}

\def\aap{A\&A}

\shorttitle{FLASH Radiation Flux--Limiter Aware Hydrodynamics}
\shortauthors{Chatzopoulos, Weide}
\begin{document}
\title
{GRAY RADIATION HYDRODYNAMICS  WITH THE {\it FLASH} CODE FOR ASTROPHYSICAL APPLICATIONS}
\author{E. Chatzopoulos\altaffilmark{1,2}, K. Weide\altaffilmark{2}}
\email{chatzopoulos@phys.lsu.edu}
\altaffiltext{1}{Department of Physics \& Astronomy, Louisiana State University, Baton Rouge, LA, 70803, USA}
\altaffiltext{2}{Department of Astronomy \& Astrophysics, Flash Center for Computational Science, University of Chicago, Chicago, IL, 60637, USA}

\begin{abstract}
We present the newly--incorporated gray radiation hydrodynamics capabilities of the {\it FLASH} code
based on a radiation flux--limiter aware hydrodynamics numerical implementation
designed specifically for applications in astrophysical problems. 
The implemented numerical methods consist of changes in the unsplit hydrodynamics 
solver and adjustments in the flux--limited radiation diffusion unit. 
Our approach can handle problems in both the strong and weak radiation--matter coupling limits as
well as transitions between the two regimes. Appropriate extensions in the ``Helmholtz'' equation
of state are implemented to treat two--temperature astrophysical plasmas involving the interaction
between radiation and matter and the addition of a new opacity unit based on the {\it OPAL} opacity database, 
commonly used for astrophysical fluids. 
A set of radiation--hydrodynamics test problems is presented aiming to showcase the new capabilities
of {\it FLASH} and to provide direct comparison to other similar software instruments available in the literature.
To illustrate the capacity of {\it FLASH} to simulate phenomena occurring in stellar explosions, 
such as shock break--out, radiative precursors and supernova ejecta heating due to the decays of radioactive 
$^{56}$Ni and $^{56}$Co, we also present 1D supernova simulations and compare the computed
lightcurves to those of the {\it SNEC} code. 
The latest public release of {\it FLASH} with these enhanced capabilities is
available for download and use by the broader astrophysics community.
\end{abstract}

\vskip 0.57 in

\section{INTRODUCTION}\label{intro}

The analysis and interpretation of electromagnetic signals is by far the main source of information
used to study astrophysical phenomena. In this regard, the importance to understand the interaction
between radiation and matter and the physics of radiation transfer is pivotal to gaining comprehensive
insights about the underlying physical mechanisms. 

Due to the complexity of radiation transport physics combined with the dynamics of strongly ionized plasmas
that can, in some cases, possess supersonic motions, most astrophysical problems
require numerical simulations for proper examination.
A number of codes have been designed that use a multitude of numerical techniques to calculate model light--curves (LCs), spectra,
polarization spectra and radiation--driven hydrodynamic flows for direct comparison with observations.

To model the diffusion of light through expanding matter for the purposes of computing supernova (SN) LCs, there are
codes that use multi--group time--dependent non--equilibrium radiative transfer (for example, the {\it STELLA} code
of \citealt{1998ApJ...496..454B}, that incorporates a radiation intensity moments scheme). Frequently, there are simpler numerical
approaches used that are based on the flux--limited diffusion approximation (FLD; \citealt{1978JQSRT..20..541M,1981ApJ...248..321L,1996ApJ...457..291C}).
Examples of such codes that are often used to compute SN LCs include the {\it SPECTRUM} code \citep{2013ApJS..204...16F},
and the publicly available {\it SNEC} code \citep{2015ApJ...814...63M}.

The radiation diffusion approximation is useful in providing us with the general emission properties and model LCs for SNe, but
a more rigorous approach requires accurate, time--dependent spectroscopic modeling. Spectroscopic modeling can be computationally
expensive, especially in 2D and 3D geometries, because it involves making use of large databases of line opacities in order to calculate emission and absorption line
profiles taking into account many factors including material composition, density, temperature and velocity. 
Currently, many spectral modeling codes are used in a post--processing manner;
pure or radiation hydrodynamical ``snapshot'' profiles are extracted from other codes and then used as inputs
to the (usually) Lagrangian grids of radiation transport codes yielding model spectra. Some spectral sythesis codes employ Monte Carlo
techniques to model radiation transfer and are optimized for both the local (LTE) and non--local thermal equillibrium (nLTE) limits. 
Examples of some of the most popular codes used include {\it CMFGEN} \citep{2012MNRAS.424..252H}, {\it SEDONA} \citep{2006ApJ...651..366K},
{\it PHOENIX} \citep{1999JCoAM.109...41H,1992JQSRT..47..433H,2004A&A...417..317H,2012ApJ...756...31V}, 
{\it SuperNu} \citep{2013ApJS..209...36W} and the open--source {\it CLOUDY} \citep{1998PASP..110..761F}
and {\it TARDIS} \citep{2014MNRAS.440..387K} codes. Some of these codes have been routinely used to study emission from expanding SN photopsheres
and have been succesfully compared to a lot of observations. 

Radiation hydrodynamics \citep{1984oup..book.....M,2007rahy.book.....C} is necessary to study the propagation and properties
of radiative shocks, supernova remnant (SNR) emission, supernova (SN) shock breakout, and radiation--driven mass loss from massive stars
near the Eddington limit, to name just a few phenomena. The applicability of the concepts of radiation hydrodynamics in sensitive fields like
nuclear weapons simulations and high-energy-density laser experiments has led to the development of codes with such capabilities in
government laboratories like the Los Alamos National Laboratory and the Lawrence Berkeley National Laboratory,
several of which are inaccessible for use by most academic researchers.

However, the advent of open--source or publicly available computational astrophysics codes like {\it MESA}
\citep{2011ApJS..192....3P,2013ApJS..208....4P,2015ApJS..220...15P} 
for stellar evolution, {\it FLASH} for hydrodynamics, {\it SNEC} for equilibrium--diffusion radiation transport
and {\it TARDIS} for spectral synthesis has energized the field of computational astrophysics
by making these essential modeling tools available for use to everyone in the community,
from graduate students to senior researchers, and thus fostering collaboration and transparency. 
Other notable examples of open--access radiation hydrodynamics codes include {\it ZEUS} \citep{1992ApJS...80..819S},
{\it HERACLES} \citep{2007A&A...464..429G}, {\it RAGE} \citep{2008CS&D....1a5005G}, {\it CRASH} \citep{2011ApJS..194...23V},
{\it RAMSES} \citep{2011A&A...529A..35C}, {\it ENZO} \citep{2011MNRAS.414.3458W} and {\it CASTRO} \citep{2011ApJS..196...20Z,2013ApJS..204....7Z}.

The {\it FLASH} \citep{2000ApJS..131..273F,2012ApJS..201...27D} adaptive--mesh refinement magnetohydrodynamics (MHD)
code is very popular amongst the numerical astrophysics community\footnote{http://flash.uchicago.edu/site/publications/flash\_pubs.shtml} --
especially in the supernova field -- with applications ranging from studies of Type Ia SNe \citep{Calder2004,Townsley2007a},
core--collapse SNe \citep{couch2013dependence,couch2013high}, pair--instability SNe \citep{chatzopoulos2013multi} and
pre--SN convection \citep{couch2013impact,chatzopoulos2014characterizing}. In addition, {\it FLASH} is amongst the best documented
software instruments online with continuous development and support provided through an active mailing list.
Nonetheless, the important component of a two--temperature (2T) radiation hydrodynamics treatment
was missing from the code thus restricting the capacity to simulate a variety of
interesting problems and obtain predictions, such as numerical SN LCs, that can be directly compared with observations.
For this reason, and to contribute to the open computational astrophysics community,
we introduce our recently implemented gray FLD radiation hydrodynamics scheme of
the {\it FLASH} code optimized for astrophysical applications and designed with emphasis on simulating physical processes that
are important within the supernova field: the Radiation Flux--Limiter Aware Hydrodynamics scheme ({\it RadFLAH}).
Our approach and numerical methods are tested in a variety of contexts and physical
domains and benchmarked against analytical predictions and published results of other codes.
The latest release of {\it FLASH} (version 4.5) includes {\it RadFLAH} and is available for
download. Some documentation is also available within the {\it FLASH} user's guide.

The paper is organized as follows. In \S~\ref{radhydro} we present the set of radiation--hydrodynamics equations
in the gray FLD limit that we are numerically solving. In \S~\ref{num_methods} we discuss in more detail
the numerical techniques implemented in the {\it FLASH} framework to solve that system of equations, namely our radiation flux--limiter
aware hydrodynamics ({\it RadFLAH}) method. A set of
test problems illustrating the new capabilities of the code is presented in \S~\ref{Test_problems}, and a special application
for 1D spherical supernova explosions is discussed in \S~\ref{SN1D}. 
Finally, in \S~\ref{Disc} we discuss our conclusions and the importance of having an open--source tool to study
radiation--hydrodynamics in astrophysics.

\section{RADIATION HYDRODYNAMICS IN THE FLUX-LIMITED DIFFUSION LIMIT}\label{radhydro}

Our implementation is based on gray FLD methods that are suitable in avoiding the main issue
of faster--than--light signal propagation when the diffusion equation is applied in the optically--thin
regime. Although FLD is one of the most commonly--used and well--established methods
\citep{1978JQSRT..20..541M,1981ApJ...248..321L,1996ApJ...457..291C} it has known limitations such as the treatment of
radiation flows in the free--streaming limit. In this regime various implementations of FLD rely on different
forms of flux--limiters that often result in notably different results when simulating standard radiation hydrodynamics
test problems (see, for example~\ref{critshocks}).

As a starting point, we take the equations for mixed-frame FLD radiation hydrodynamics
developed in \citet{2007ApJ...667..626K}. Adopting notation for our purposes, we write
\begin{eqnarray}
\frac{\partial \rho}{\partial t} + \nabla \cdot (\rho \vv ) = 0\,,\,\label{eq:Continuity}\\
\frac{\partial (\rho \vv )}{\partial t} + \nabla \cdot (\rho \vv \mathop\otimes\mathop \vv ) + \nabla \pmat  + \lambda \nabla \Erad  = 0\,,\,\label{eq:Momentum}\\
\nonumber\\
\frac{\partial \Emat }{\partial t} + \nabla \cdot\left[(\Emat  + \pmat ) \vv \right] 
- \lambda \left( 2 \frac{\kappaP }{\kappaR }-1 \right) \vv  \cdot \nabla \Erad  \nonumber \\
= - \kappaP (4 \pi B - c\Erad )
\,,\,\label{eq:MatEnergy_den}\\
\nonumber\\
\frac{\partial \Erad }{\partial t} 
+ \nabla \cdot \left[(1+\lambda') \Erad \vv  \right]
+\lambda \left(2 \frac{\kappaP }{\kappaR } - 1 \right) \vv  \cdot \nabla \Erad  
\nonumber \\
= \nabla \cdot \left(\frac{c \lambda}{\kappaR } \nabla \Erad  \right) + \kappaP  (4 \pi B - c\Erad ) 
\,.\,\label{eq:RadEnergy}
\end{eqnarray}
Here $\kappaP$ and $\kappaR$ are the Planck (absorption) and Rosseland (transport) coefficients respectively and
$B$ is the Planck function.
Also, $\Emat$ is the 
matter energy density, 
defined by the relation $\Emat=\rho\emat + \rho\frac{\vv^2}{2}$ (where $\emat$ is
specific internal matter energy), and
$\Erad$ is the radiation energy density.
We make the approximation that the flux limiter $\lambda$ depends on radiation energy density $\Erad$
in the lab frame (rather than a comoving density $\Erad^{(0)}$).
Thus  $\lambda=\lambda(R)$ depends on 
the quantity
$R=\frac{\left|\nabla \Erad  \right|}{\kappa_R \Erad}$, and we have
further introduced the abbreviation
$\lambda' = \frac{1 - f}{2}$, where
$f=\lambda + \lambda^2 R^2$ is the Eddington factor.
Note that both $\lambda$ and $\lambda'$ have similar asympotic behavior for both
the diffusion limit ($\lambda,\lambda' \rightarrow 1/3$ for $R\rightarrow 0$)
and 
the free--streaming limit ($\lambda,\lambda' \rightarrow 0$ for $R\rightarrow \infty$);
moreover, as pointed out in \citet{2011ApJS..196...20Z}; their difference remains small
for all $0<R<\infty$.

Our implementation uses operator splitting to separate this system of equations
into an ``enhanced hydro'' subsystem and a ``radiation transfer'' subsystem.
The latter describes the effect of the terms written on the
right--hand side in Equations~\ref{eq:Continuity}--~\ref{eq:RadEnergy} above, and is equivalent
to
\begin{eqnarray}
\rho\frac{\partial \emat }{\partial t}
&=& - \kappaP (4 \pi B - c\Erad )
\,,\,\label{eq:MatEnergy_den_RadTrans}\\
\nonumber\\
\frac{\partial \Erad }{\partial t} 
&=& \nabla \cdot \left(\frac{c \lambda}{\kappaR } \nabla \Erad  \right) + \kappaP  (4 \pi B - c\Erad ) 
\,.\,\label{eq:RadEnergy_RadTrans}
\end{eqnarray}
The former 
consists of
Equations.~\ref{eq:Continuity}--\ref{eq:RadEnergy} with right hand sides set to 0;
we call our approach to solving this system Radiation Flux--Limiter Aware Hydrodynamics ({\it RadFLAH}).
By adding the last two of those modified equations, 
\begin{eqnarray}
\frac{\partial \Emat }{\partial t} + \nabla \cdot[(\Emat  + \pmat ) \vv]
- \lambda ( 2 \frac{\kappaP }{\kappaR }-1) \vv  \cdot \nabla \Erad
= 0
\,,\,\label{eq:MatEnergy_den_0}\\
\nonumber\\
\frac{\partial \Erad }{\partial t} 
+ \nabla \cdot [(1+\lambda') \Erad \vv  ]
+\lambda (2 \frac{\kappaP }{\kappaR } - 1 ) \vv  \cdot \nabla \Erad  
= 0
\,,\,\label{eq:RadEnergy_0}
\end{eqnarray}
we get the following equation:
\begin{eqnarray}
\frac{\partial}{\partial t}( \Emat+\Erad ) 
+ \nabla \cdot\left[(\Emat  + \pmat 
+
(1+\lambda') \Erad )\vv  \right]
= 0
\,.\,\label{eq:TotEnergy}
\end{eqnarray}
This can also be written
\begin{eqnarray}
\frac{\partial \Etot }{\partial t} 
+ \nabla \cdot\left[(\Etot  + \ptot +\pcorr)\vv  \right]
= 0
\,,\,\label{eq:TotEnergy_prime}
\end{eqnarray}
with $\Etot= \Emat + \Erad $ and $\ptot= \pmat + \lambda \Erad $
and a small correction term $\pcorr = ( \lambda' - \lambda) \Erad$.

For further reference, we also write an equivalent equation for matter internal specific energy:
\begin{eqnarray}
\frac{\partial (\rho \emat )}{\partial t} + \nabla \cdot(\rho \emat  \vv )
+ \pmat \nabla\cdot\vv
- 2 \lambda  \frac{\kappaP }{\kappaR }   \vv  \cdot \nabla \Erad
= 0 
\,.\,\label{eq:MatIntEnergy} \nonumber \\
\end{eqnarray}

\section{NUMERICAL METHODS}\label{num_methods}

The goal of the {\it RadFLAH} code is to solve the (overdetermined)
system of five equations \eqref{eq:Continuity},\eqref{eq:Momentum},\eqref{eq:MatEnergy_den_0},\eqref{eq:RadEnergy_0},\eqref{eq:TotEnergy_prime}.
This could be done by directly implementing a hyperbolic solver for a system 
consisting of equations \eqref{eq:Continuity}, \eqref{eq:Momentum}, and any two of \eqref{eq:MatEnergy_den_0}, \eqref{eq:RadEnergy_0}, and \eqref{eq:TotEnergy_prime}.
We will instead first solve the system of three equations \eqref{eq:Continuity},\eqref{eq:Momentum},\eqref{eq:TotEnergy_prime}
numerically for a time step,
thus computing new values of $\rho$, $\vv$, and total energy $\Etot$,
and then use this solution together with \eqref{eq:MatIntEnergy} and \eqref{eq:RadEnergy_0} to distribute the
total energy change (computed directly from \eqref{eq:TotEnergy_prime}) to the energies $\Emat$ and $\Erad$.

{\it FLASH} already provides a variety of directionally unsplit methods for solving the system of Euler equations of
hydrodynamics (HD), as well as the equations of magnetohydrodynamics (MHD).
These are based on the Godunov approach and feature a variety of
Riemann solvers, orders of reconstruction, slope limiters, and related features.
The HD and MHD solvers can work with a variety of equation of state (EOS)
models by using a formulation derived from \cite{ColellaGlaz}.
In addition to advancing the core variables of HD or MHD, 
{\it FLASH} can also advect arbitrary additional variables $X$ (``mass scalars''),
equivalent to solving additional equations
\begin{equation}
\frac{\partial (\rho X)}{\partial t} + \nabla \cdot (\rho X \vv ) = 0\,.\,\label{eq:MassScalar}
\end{equation}
Our approach has been to reuse as much of this existing code as possible.
Here we outline this approach; some more implementation details can be found in the appendix.

First, we write the fluid state in conservative form as
\begin{equation}
{\mathitbf{U}} = \begin{pmatrix}
\rho\cr
\rho\vv\cr
\Etot\cr
\rho \emat\cr
\Erad\cr
X_1\rho\cr
\vdots\cr
X_n\rho
\end{pmatrix}
\end{equation}
and our evolution equations as
\begin{multline}
\frac{\partial}{\partial t}{\mathitbf{U}} = f^{[1]} + f^{[2]} + f^{[3]} + f^{[4]} \\ =  f_{hyperbolic}+f_{fixup}+f_{Lorentz}+f_{transp}.
\end{multline}

Here
\begin{equation}
f^{[1]} = f_{hyperbolic} =
\begin{pmatrix}
-\nabla\cdot(\rho\vv)\cr
-\nabla\cdot(\rho\vv\vv) - \nabla \pmat - \lambda \nabla \Erad \cr
-\nabla\cdot \left[
(\Etot+\ptot + \pcorr)\vv
\right]
\cr
-\nabla\cdot \left(
\rho \emat \vv
\right) \cr
-\nabla\cdot \left[
(1+\lambda')\Erad\vv
\right] \cr
-\nabla\cdot(\rho X_1 \vv)\cr
\vdots\cr
-\nabla\cdot(\rho X_n \vv)\cr
\end{pmatrix},
\end{equation}
\begin{equation}
f^{[2]} = f_{fixup} =
\begin{pmatrix}
0\cr
\mathbf{0}\cr
0 
\cr
-\pmat\,\nabla\cdot \vv
\cr
\lambda\vv\cdot
\nabla\Erad
\cr
0\cr
\vdots\cr
0
\end{pmatrix},
\end{equation}
\begin{equation}
f^{[3]} = f_{Lorentz} =
\begin{pmatrix}
0\cr
\mathbf{0}\cr
0\cr
2 \lambda  \frac{\kappaP }{\kappaR }   \vv  \cdot \nabla \Erad
\cr
- 2 \lambda  \frac{\kappaP }{\kappaR }   \vv  \cdot \nabla \Erad
\cr
0\cr
\vdots\cr
0
\end{pmatrix},\label{eq:fLorentz}
\end{equation}
and
\begin{equation}
f^{[4]} = f_{transp} =
\begin{pmatrix}
0\cr
\mathbf{0}\cr
\nabla \cdot \left(\frac{c \lambda}{\kappaR } \nabla \Erad  \right) 
\cr
 - \kappaP  (4 \pi B - c\Erad ) 
\cr
\nabla \cdot \left(\frac{c \lambda}{\kappaR } \nabla \Erad  \right) + \kappaP  (4 \pi B - c\Erad ) 
\cr
0\cr
\vdots\cr
0
\end{pmatrix}.\label{eq:ftransp}
\end{equation}

The numerical advance of the solution from state $U^{(n)}$ to $U^{(n+1)}$ by a 
time step $\Delta t$ can then be performed in several successive
phases $\rm p \in \{1,2,3,4\}$: 

\begin{equation}
U^{(n)[1]} = U^{(n)}       + \Delta t \\  f^{[1]},
\end{equation}
\begin{equation}
U^{(n)[p]} = U^{(n)[\rm p -1]}       + \Delta t \\  f^{[\rm p]}, \\\\\ \rm p = 2,...,4,
\end{equation}
\begin{equation}
U^{(n+1)} = U^{(n)[4]},
\end{equation}

where the term $f^{[1]}$ corresponds to divergence of fluxes while
the other terms $f^{[2,3,4]}$ are not. We now briefly
describe the meaning of each term:
\begin{itemize}
\item{$U^{(n)[1]}$: This term corresponds to the conservative form of our modified hydrodynamics 
implementation that is described in detail below (Section~\ref{mod_hydro}).}
\item{$U^{(n)[2]}$: Non--hyperbolic additional modified--hydro term. In practice, these steps are computed within the regular 
{\it FLASH} unsplit hydrodynamics unit, as an add--on action after the main update. 
Note, that this term only modifies the component energies, not conserved totals.}
\item{$U^{(n)[3]}$:  An additional coupling term of relativistic nature. This term is also computed within the hydrodynamics unit, 
as an add--on action after the main update; but could also be separated out of hydro and be done as part of phase 4 
(we plan to include this capability in a future version). This term only modifies the individual (radiation, matter) component energies, 
not the conserved totals.}
\item{$U^{(n)[4]}$: The radiation transport component. This is completely separate from the hydrodynamics component and
is included here to facilitate term--by--term comparison with other papers.}
\end{itemize}
We must add that the implementation of $U^{(n)[4]}$ is not the central subject of this paper, since we are using two 
pre--existing methods of previous versions of {\it FLASH}. What is new is that we are using them in the context
of the 2T {\it RadFLAH} implementation. These original methods include a flux limiter and allow us to expand
to alternative flux limiter implementations. As such, we are using the same flux--limiter formulation for additional purposes 
within the modified hydrodynamics implementation.

\subsection{{\it Modified hydrodynamics.}}\label{mod_hydro}

The system \eqref{eq:Continuity},\eqref{eq:Momentum},\eqref{eq:TotEnergy_prime} to be solved
already looks like the Euler system {\it FLASH} can solve, for a fluid consisting of
matter and radiation components, with just a few differences:
\begin{enumerate}
\item The  momentum equation \eqref{eq:Momentum} 
contains a term $\lambda \nabla \Erad$ (instead of  $\nabla\lambda\Erad$;
a non--flux limiter--aware hydro formulation would have the term $\nabla\frac13\Erad$ here).

We account for this by advecting additional information from which (for, e.g., the $i$--drection) the radiation energy ${\Erad}_{i \pm 1/2,j,k}$ at cell
interfaces can be reconstructed, and then computing
$\displaystyle  \lambda_{i,j,k} \frac{{\Erad}_{i + 1/2,j,k}   -   {\Erad}_{i - 1/2,j,k}   } {2}   $
using $\lambda$ values computed from the previous solution state.
\item The pressure of the radiation field in the $\ptot$ term of the energy equation \eqref{eq:TotEnergy_prime}
is reduced to an effective pressure $\Pradeff = \lambda \Erad$ by scaling with $3\lambda$.
(A non--flux limiter--aware hydro formulation would have $\Prad = \frac13\Erad$.)

We account for this by replacing $\Prad$ by $\Pradeff$ in the state that is fed to the hydro solver
for reconstruction, flux compuyation, and  updating of conservative variables.
\item \label{item3}The difference between $\lambda$ and $\lambda'$ leads to the  $\pcorr$ term of energy equation \eqref{eq:TotEnergy_prime}.
We navigate this by advecting a correction and adding it to the fluxes for the energy equation.
\end{enumerate}

\subsubsection{Flux computation}
Following \citet{2011ApJS..196...20Z} on the gray radiation hydrodynamics implementation in the {\it CASTRO} code, 
we note $\lambda\approx\lambda'$ in particular for the \citet{1981ApJ...248..321L} (LP) flux limiter;
we assume in the following that this approximate equality holds true for the flux limiter used.
The Godunov method ultimately involves computing fluxes
by solving 1D Riemann problems at cell interfaces.
Each Riemann problem yields a solution consisting of a ``fan'' made up of
several waves; the number of waves is determined by the number
of distinct eigenvalues of a Jacobian matrix of the form:
$$\begin{pmatrix}
v&\rho&0&0\cr 0&v&{{1}\over{\rho}}&{{\lambda}\over{\rho}}
 \cr 0&\gamma\,p&v&\left(1-\gamma\right)\,v\,K\,\lambda\cr 0&
 \left(\lambda+1\right)\Erad&0&v\,\left(K\,\lambda+1\right)\cr \end{pmatrix}$$
derived from the equations, where 
$\gamma$ is an effective adiabatic index of the matter
that determines the matter--only sound speed,
and we use the abbreviation $K= \frac{\kappaP }{\kappaR } $.

As shown in \citet{2011ApJS..196...20Z}, the set of eigenvalues for a full hyperbolic system,
say \eqref{eq:Continuity},\eqref{eq:Momentum},\eqref{eq:MatEnergy_den_0},\eqref{eq:RadEnergy_0},
degenerates to the smaller set of eigenvalues of our system \eqref{eq:Continuity},\eqref{eq:Momentum},\eqref{eq:TotEnergy_prime} 
under the approximation $\lambda'=\lambda$, if we further assume $K=0$.
The eigenvalues in this case, $u-c_s,u,u+c_s$ (where $u$ is a velocity component
normal to the cell face for which a Riemann problem is solved), depend on the modified sound speed
\begin{equation}
c_s = \sqrt{\gamma\frac{p}{\rho}+(1+\lambda)\frac{\Pradeff}{\rho}}
\end{equation}
We note that this is the same sound speed we get with {\it FLASH} for
a fluid composed of matter and (appropriately scaled) radiation.

\subsection{{\it Flux-limited diffusion solver.}}\label{flux_lim}

We are using the FLD solver already available in previous
versions of FLASH. While the default implementation provides for radiation transport 
in multiple energy groups, we do not yet make use of this multigroup feature
for {\it RadFLAH} applications.

In addition to this default multigroup implementation, FLASH also includes
an iterative solver for strong radiation--matter coupling\label{exp_relax}
as an experimental alternative ({\tt ExpRelax}). This is a module within the \code{RadTrans} unit
and is based on the {\it RAGE} code paper \citep{2008CS&D....1a5005G}. {\tt ExpRelax}
can handle the coupling of energy and radiation at high temperatures via
an exponential relaxation method resulting in better accuracy, larger timesteps and therefore reduced computing time.
The exponential differencing of the material energy equation is useful in a class of problems in which radiation floods a
region of space and serves to heat a contained body, and allows a smooth transition to equilibrium diffusion.

\subsection{{\it Extended 2T Helmholtz Equation of State}}\label{2T_helm}

In general, the EOS is implemented as a subroutine that, given a set of
variables describing the fluid state at a physical location, updates
some of them as functions of some others, ensuring that the resulting
set of values represents a consistent state.
To be generally usable to the rest of the code, The EOS routine must
be callable in several modes, which differ by which variables are considered
as the independent (input) ones: at least, a mode in which temperatures
are inputs ("{\tt dens\_temp}") and another one in which energy variables are
inputs ("{\tt dens\_ei}") are required. Additionally there is the question of the ``number of temperatures''.
In the standard hydrodynamics version of {\it FLASH}, a one--temperature model (1T) is assumed.
The EOS then simply provides $E_{\rm m} (T_{\rm m})$, and $T_{\rm m} (E_{\rm m})$.

A configuration variant available since {\it FLASH} version 4.0 tailored
for high--energy density physics (HEDP) applications uses a three--temperature
model (3T), with separate state variables -- temperatures, energies, and also pressures --
for three separate components (ions (``i''), electrons (``e'') and radiation (``r'')).
The EOS routine then provides $E_{\rm i} (T_{\rm i})$, $E_{\rm e} (T_{\rm e})$, $E_{\rm r} (T_{\rm r})$
and $T_{\rm i} (E_{\rm i})$, $T_{\rm e} (E_{\rm e})$, $T_{\rm r} (E_{\rm r})$. 

For the current work, in which we want to represent two separate
components, we have created another variant  of the \code{Eos} interface.
We refer to this approach as 2T(M+R). The EOS routine provides
$E_{\rm m} (T)$, $T(E_{\rm m})$, $E_{\rm r} (T_{\rm r})$ and $T_{\rm r} (E_{\rm r})$ in this case.
While the last two equations (for the radiation component)
have a rather simple implemention given by Planck's law
and could be easily handled completely outside of the EOS code unit
(leaving the latter to deal exclusively with ``matter''),
we have chosen not to do so; this is for practical purposes (minimization of interface changes),
to emphasize the continuity with configurations of {\it FLASH} in 1T and 3T modes,
and to avoid introducing knowledge of radiation physics into parts of the code that are so far ignorant thereof.

This new implementation is based on existing {\it FLASH} code capabilities for 3T EOS models that deal with three
independent components (ions, electrons, radiation) of input and output variables, modified
to now act on two independent components (matter and radiation).
The variable slot previously used for electrons is reinterpreted to stand for matter,
while the slot for ions is ignored.
In particular, we have created a 2T variant of the Helmholtz EOS implementation
described in \citep{2000ApJS..131..273F} and in the {\it FLASH} users guide.
We emphasize that what is new here is merely the interface provided
by the EOS unit to other parts of the code.
The underlying lower--level code, including the essential code and tables used
for interpolating the Helmholtz free energy of the electron component,
are still the same as in 1T {\it FLASH}. 

In addition, some changes were made to make the Helmholtz EOS more robust:
when called with a $T<10^4\,\mathrm{K}$, the table--based values are extented
according to ideal--gas law.

\subsection{{\it Summary of Code Changes}}
A summary of additions and changes to the FLASH code that were implemented
as part of this work:
\begin{itemize}
\item Modified Hydro:
\begin{itemize}
  \item Made ``flux--limiter aware'' by implementing additional terms
        described in this paper
  \item Optional spatial smoothing of flux limiter variable in Hydro. {\bf In gathering practical experience with the method as described, we found 
that the addition of flux--limiter dependent terms to the hyperbolic system sometimes lead to strong oscillatory behavior of the solution 
in some locations (usually in the low--density gas regions). We found that applying one or more passes of a simple 3-point smoother to 
the discrete grid representation of the flux limiter would remedy such unstable behavior.}
\end{itemize}
\item 2T (M+R) Helmholtz equation of state.
\item Improved Eos robustness.
\item \code{OUTSTREAM} boundary for free--streaming radation conditions
{\bf at the outer boundary of a spherical domain}.
\item Added Opacity implementation that uses OPAL tables \citep{1996ApJ...464..943I}.
\end{itemize}

\section{TEST PROBLEMS}\label{Test_problems}

The following problems aim to test the newly implemented {\it RadFLAH} method in
{\it FLASH} as described in the previous sections. All test problem simulations are done in a 1D spherical
grid (except the shock--tube test problem (Section~\ref{ShockTube}) in 1D Cartesian geometry), 
and the main simulation parameters (domain size, simulation time, resolution, opacities and boundary
conditions) are summarized in Table~\ref{T1}. For all tests, the \citet{1981ApJ...248..321L} (LP) flux--limiter
is used. Aside from testing the newly implemented {\it FLASH} capabilities, we choose our test simulation
parameters in a way that we can directly benchmark our results against those of other codes and available analytical results, 
namely the ones presented by \citet{2007ApJ...667..626K} and {\it CASTRO} \citep{2011ApJS..196...20Z}, among others.

\setcounter{table}{0}
\begin{deluxetable*}{lcccccccccc}
\tablewidth{0pt}
\tablecaption{Simulation parameters for the {\it RadFLAH} test problems.}
\tablehead{
\colhead {Test problem (\S)} &
\colhead {$\Delta R$~(cm)} &
\colhead{$t_{\rm sim}$~(sec)} &
\colhead {CFL} &
\colhead {$\Delta r_{\rm min}$~(cm)} &
\colhead {$\kappa_{\rm R}$~(cm$^{-1}$)$\dagger$} &
\colhead {$\kappa_{\rm P}$~(cm$^{-1}$)$\dagger$} &
\colhead {BC$_{\rm hydro}$~(inner/outer)} &
\colhead {BC$_{\rm rad}$~(inner/outer)} &
\\}
\startdata
\S~\ref{Thermal_Eq}   & 1.0 & $10^{-5}$ & 0.8 & 0.1 & $4.0 \times 10^{-8}$ & $4.0 \times 10^{-8}$ & {\tt reflect} & {\tt reflecting} \\
&  & & & & & & {\tt reflect} & {\tt reflecting} \\
\S~\ref{Marshak}   & 20.0 & $10^{-10}$ & 0.8 & $6.77 \times 10^{-2}$ & 1.0 & 1.0 & {\tt reflect} & {\tt marshak} \\
&  & & & & & & {\tt outflow} & {\tt outflow} \\
\S~\ref{Lowrie}   & 0.06 & $2 \times 10^{-11}$ & 0.8 & $5.0 \times 10^{-5}$ & 788.03 & 422.99 & {\tt outflow} & {\tt outflow} \\
&  & & & & & & {\tt outflow} & {\tt outflow} \\
\S~\ref{critshocks}   & $7.0 \times 10^{10}$ & $5.80 \times 10^{4}$ & 0.8 & $1.38 \times 10^{8}$ & $3.12 \times 10^{-10}$ & $3.12 \times 10^{-10}$ & {\tt reflect} & {\tt reflecting} \\
&  & & & & & & {\tt outflow} & {\tt vacuum} \\
\S~\ref{Rad_opt_thin}$\dagger$   & $10^{12}$ & $10^{6}$ & 0.5 & $1.95 \times 10^{9}$ & $4 \times 10^{-6}$ & $4.0 \times 10^{-10}$ & {\tt reflect} & {\tt reflecting} \\
&  & & & & & & {\tt extrapolate} & {\tt outstream} \\
\S~\ref{Bondi}$\dagger$               & $2.50 \times 10^{13}$ & $1.54 \times 10^{7}$ & 0.8 & $9.77 \times 10^{10}$ & 0.4 & 0.0 & {\tt user} & {\tt dirichlet} \\
&  & & & & & & {\tt user} & {\tt dirichlet} \\
\S~\ref{ShockTube}                       & 100.0 & $10^{-6}$ & 0.8 & 0.78 & $1 \times 10^{8}$ & $1 \times 10^{6}$ & {\tt outflow} & {\tt vacuum} \\
&  & & & & & & {\tt outflow} & {\tt vacuum} \\
\S~\ref{RadShock}  -- Case 1             & $10^{14}$ & $10^{6}$ & 0.6 & $9.766 \times 10^{10}$ & $2.0 \times 10^{-10}$ & $2.0 \times 10^{-16}$ & {\tt reflect} & {\tt vacuum} \\
&  & & & & & & {\tt outflow} & {\tt vacuum} \\
\S~\ref{RadShock} -- Case 2              & $10^{14}$ & $10^{6}$ & 0.6 & $9.766 \times 10^{10}$ & $2.0 \times 10^{-10}$ & $2.0 \times 10^{-7}$ & {\tt reflect} & {\tt vacuum} \\
&  & & & & & & {\tt outflow} & {\tt vacuum} \\
\enddata 
\tablecomments{Where $\Delta R$ is the size of the computational domain (in 1D spherical coordinates), $t_{\rm sim}$ the total simulation time,
CFL the CFL number, $\Delta r_{\rm min}$ the maximum resolution (or minimum cell size), $\kappa_{\rm R}$ and $\kappa_{\rm P}$ the transport (Rosseland)
or absorption (Planck) mean opacity accordingly and BC$_{\rm hydro}$, BC$_{\rm rad}$ the outer boundary condition chosen for hydrodynamics and radiation
respectively. $\dagger$ The chosen input opacities for these tests are in units of cm$^{2}$~g$^{-1}$. For more details on the specifics of the chosen
boundary conditions please refer to the {\it FLASH} user guide. 
\label{T1}}
\end{deluxetable*}

\subsection{{\it Thermal equilibration }}\label{Thermal_Eq}

The first setup that we reproduce in order to test our {\it RadFLAH} implementation was introduced by \citep{2001ApJS..135...95T} and is
used to examine how accurately the code can model the approach to thermal equilibrium between radiation and matter
in a static uniform field of gas and radiation. Our simulation setup is using the same initial conditions as those used by 
\citet{2011ApJS..196...20Z}; a uniform density $\rho = 10^{-7}$~g~cm$^{-3}$, a Planck (absorption) coefficient
$\kappa_{\rm P} = 4 \times 10^{-8}$~cm$^{-1}$, a mean molecular weight $\mu =$~0.6 and an adiabatic index $\gamma =$~5/3.
{\bf The initial radiation temperature is set to $T_{\rm r} = 3.39 \times 10^{6}$~K (equivalent to radiation energy
density $E_{\rm r} = 10^{12}$~erg~cm$^{-3}$). }
A fixed timestep of $10^{-11}$~s is chosen for the simulation. We run two cases for two different choices for the 
initial internal energy density of the gas:  $10^{10}$~erg~cm$^{-3}$ {\bf (corresponding to initial gas temperature $T_{\rm m} = 4.81 \times 10^{8}$~K)} 
and 100~erg~cm$^{-3}$. Assuming that only a small fraction of the radiation energy is exchanged into gas energy, an analytic solution can be derived
by solving the ordinary differential equation:
\begin{equation}
\frac{d(\rho e)}{dt} = -c \kappa_{\rm P} \left(a T^{4} - E_{\rm r} \right).
\end{equation}

\begin{figure*}
\begin{center}
\includegraphics[angle=0,width=15cm,trim=0.in 0.25in 0.5in 0.15in,clip]{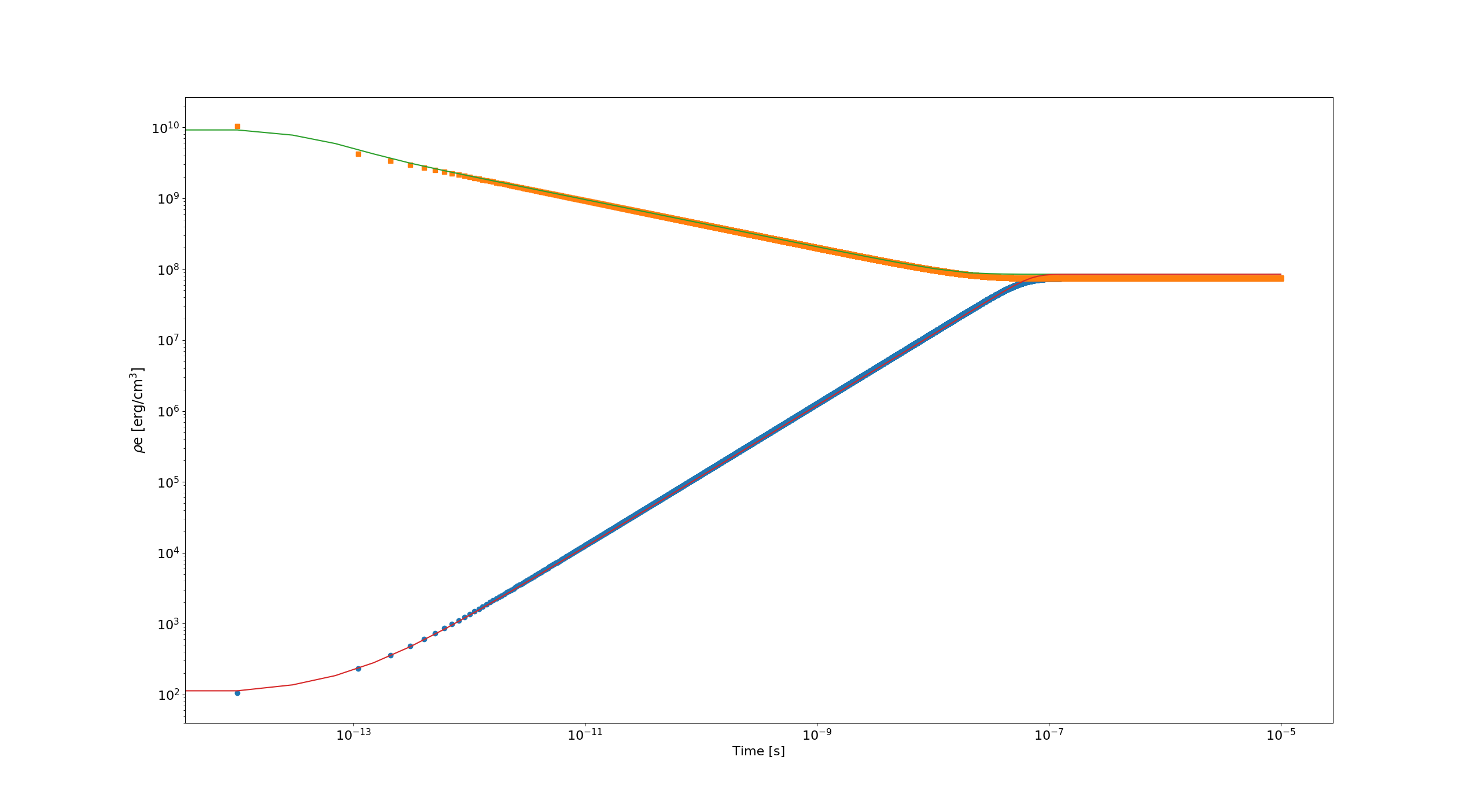}
\caption{Approach to thermal equilibrium test problem (~\ref{Thermal_Eq}). The evolution of gas internal energy density
is shown for two cases: initial gas energy density of $10^{10}$~erg~cm$^{-3}$ (upper solid curve and filled square symbols) and initial
gas energy density of 100~erg~cm$^{-3}$ (lower solid curve and filled circle symbols). Solid curves represent the results of our
test simulation with {\it RadFLAH} and symbols the analytic solutions of \citet{2001ApJS..135...95T}.}\label{Fig:energyxchange}
\end{center}
\end{figure*}

The results of our test are plotted against the analytic solution in Figure~\ref{Fig:energyxchange}. Very good agreement
is found for both choices for the initial gas energy density and in both cases, equilibration is reached in $\simeq 10^{-7}$~s.

\subsection{{\it Non--equilibrium Marshak wave}}\label{Marshak}

A useful test to evaluate the coupling between matter and radiation is the non--equilibrium Marshak wave problem. In this test
the initial setup is a simulation domain with no radiation and a static, uniform--density, zero temperature gas. An incident radiation
flux, $F_{\rm inc}$, is introduced on the left boundary of the domain (at $x = 0$) leading to the formation of a wave that progapates toward
the right boundary. Analytic solutions to the non--equilibrium Marshak wave test problem are derived by \citet{1996JQSRT..56..337S}
and can be expressed in a dimensionless form as follows \citep{1979JQSRT..21..249P} :

\begin{equation}
x^{\prime} \equiv \sqrt{3} \kappa x,
\end{equation}

\begin{equation}
\tau \equiv \left(\frac{4 a c \kappa}{\alpha}\right) t,
\end{equation}

\begin{equation}
u(x^{\prime},\tau) \equiv \left(\frac{c}{4}\right) \left(\frac{E_{\rm r} (x^{\prime},t)}{F_{\rm inc}}\right),
\end{equation}

\begin{equation}
v(x^{\prime},\tau) \equiv \left(\frac{c}{4}\right) \left(\frac{a T^{4} (x^{\prime},t)}{F_{\rm inc}}\right),
\end{equation}

where $x^{\prime}$, $\tau$, $u$ and $v$ are the dimensionless spatial coordinate, time, radiation and matter energy density accordingly and
$\alpha$ is a parameter controlling the volumetric heat capacity, and therefore the EOS of the matter: $c_{\rm V} = \alpha T^{3}$ with
$4 a/\alpha = \epsilon$. In our test run we use $\epsilon =$~0.1 and and the matter is assumed to be gray with
$\kappa_{\rm P} = \kappa_{\rm R} =$~1.0~cm$^{-1}$.

\begin{figure*}
\begin{center}
\includegraphics[angle=0,width=15cm,trim=0.in 0.25in 0.5in 0.15in,clip]{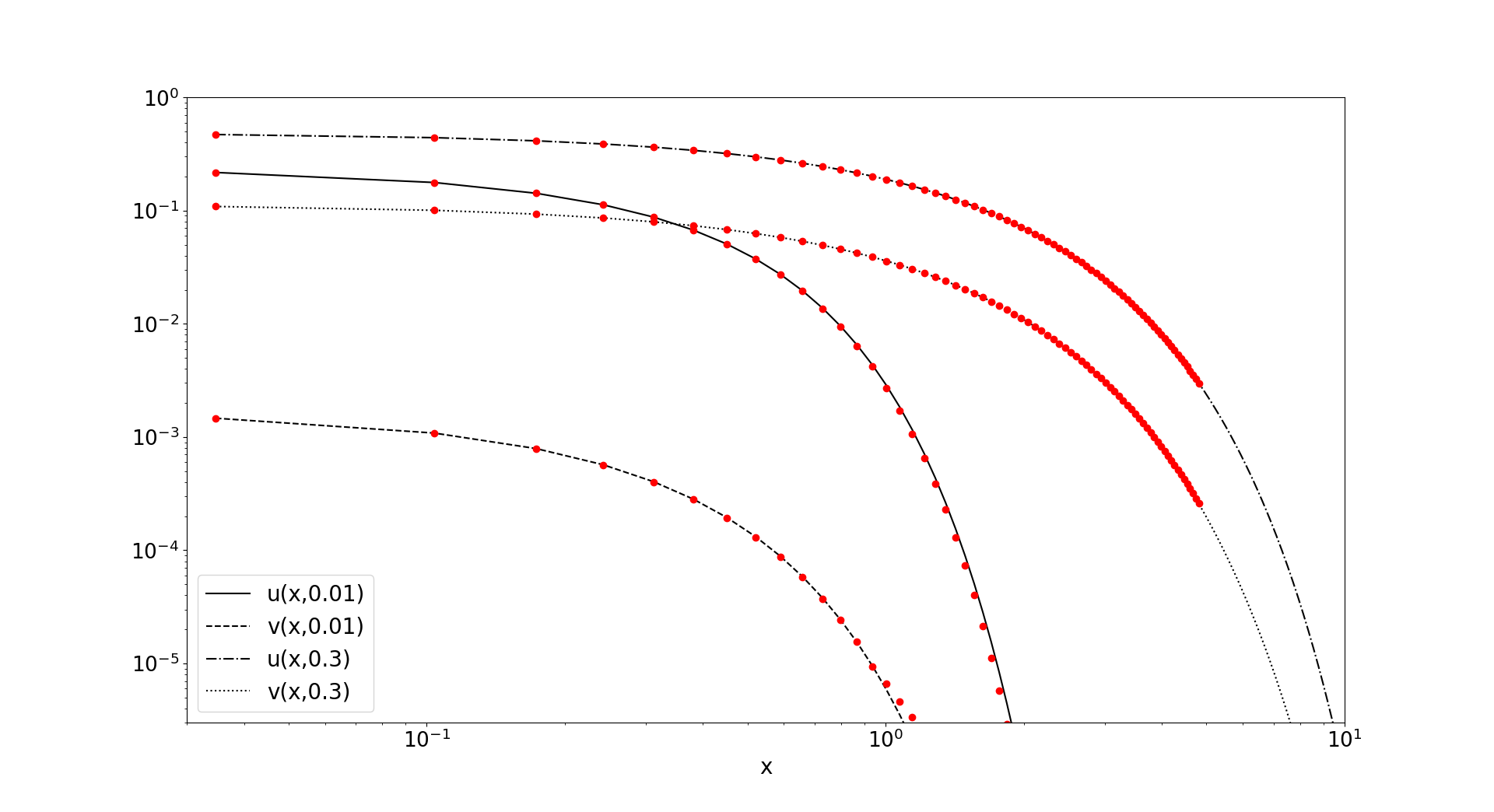}
\caption{The non--equilibrium Marshak wave test problem (\ref{Marshak}). Curves represent our numerical results
with {\it RadFLAH} while filled red circles the analytic results as described by \citet{1996JQSRT..56..337S}. Dimensionless
radiation ($u$) and matter ($v$) energy density is plotted for two choices of dimensionless time: $\tau =$~0.01
and $\tau =$~0.3.}\label{Fig:suolson}
\end{center}
\end{figure*}

In order to properly setup this test problem we had to introduce a new {\tt marshak} radiation boundary condition ($BC_{\rm rad}$) in {\it FLASH}
identical to the one represented by Equation 3 of \citet{1996JQSRT..56..337S}. This new BC is essentially a combination of the already
available {\tt vacuum} and {\tt dirichlet} BCs in the code. Figure~\ref{Fig:suolson} shows the results of our simulation in dimensionless units
for two different choices of dimensionless time ($\tau =$~0.01 and $\tau =$~0.3). Comparison with the contemporaneous analytic solutions
shows excellent agreement.

\subsection{{\it Steady radiative shock structure}}\label{Lowrie}

Another common stress--test for radiation hydrodynamics codes is that of the structure of steady radiative shocks. Radiation--matter interactions
can change the radiation and matter temperature profiles as well as the density profile of a shock. Furthermore, numerical
results for this test can be verified against semi--analytical solutions that were presented by \citet{2008ShWav..18..129L}. This evaluation test
has been used by many radiation hydrodynamics implementations \citep{2007A&A...464..429G,2011ApJS..196...20Z,2015ApJS..217....9R} thus it
is critical that we successfully reproduce it with {\it RadFLAH}.

We closely follow the initial setup described in \citet{2008ShWav..18..129L} and run this test problem for two shock strength cases: a subcritical
(``Mach 2''; $\mathcal{M} =$~2) case and a supercritical (``Mach 5''; $\mathcal{M} =$~5) case. The analytical solutions for the
shock structures (radiation, matter temperature and density) are shown in Figures 8 and 11 of \citet{2008ShWav..18..129L} respectively.
Our 1D simulation domain extends in the range $-0.03 < x < 0.03$~cm and
consists of ideal gas with $\gamma =$~5/3 and mean molecular weight $\mu =$~1.0.
The Planck and Rosseland coefficients are set to $\kappa_{\rm P} =$~422.99~cm$^{-1}$ and $\kappa_{\rm R} =$~788.03~cm$^{-1}$ respectively. A
discontinuity is placed at $x =$~0.0~cm separating the domain in left (``L'') and right (``R'') states with the following properties:

\begin{itemize}
\item{{\it Mach 2 case}: $\rho_{\rm L} =$~1.0~g~cm$^{-3}$, $T_{\rm L} =$~100~eV, $\rho_{\rm R} =$~2.286~g~cm$^{-3}$, $T_{\rm R} =$~207.756~eV.}
\item{{\it Mach 5 case}: $\rho_{\rm L} =$~1.0~g~cm$^{-3}$, $T_{\rm L} =$~100~eV, $\rho_{\rm R} =$~3.598~g~cm$^{-3}$, $T_{\rm R} =$~855.720~eV.}
\end{itemize}

\begin{figure*}
\begin{center}
\includegraphics[angle=0,width=15cm,trim=0.in 0.25in 0.5in 0.15in,clip]{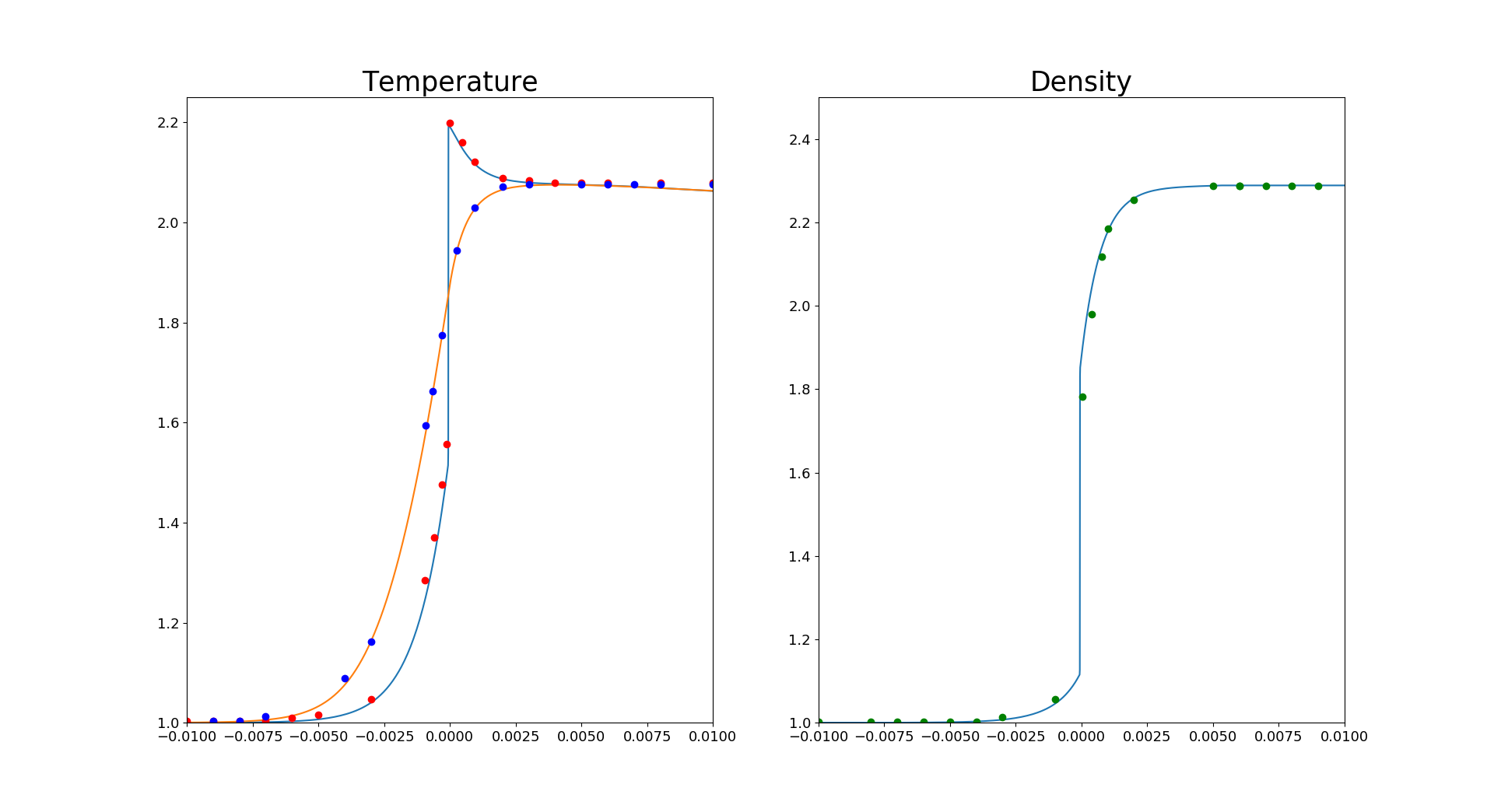}
\caption{Temperature ({\it left panel}) and density ({\it right panel}) profiles for a Mach 2 
($\mathcal{M} =$~2) subcritical radiative shock (\ref{Lowrie}). The orange and blue curves in the left panel correspond to radiation and material 
temperature respectively.The filled circles correspond to the semi--analytical results of \citet{2008ShWav..18..129L}.}\label{Fig:LEmach2}
\end{center}
\end{figure*}

\begin{figure*}
\begin{center}
\includegraphics[angle=0,width=15cm,trim=0.in 0.25in 0.5in 0.15in,clip]{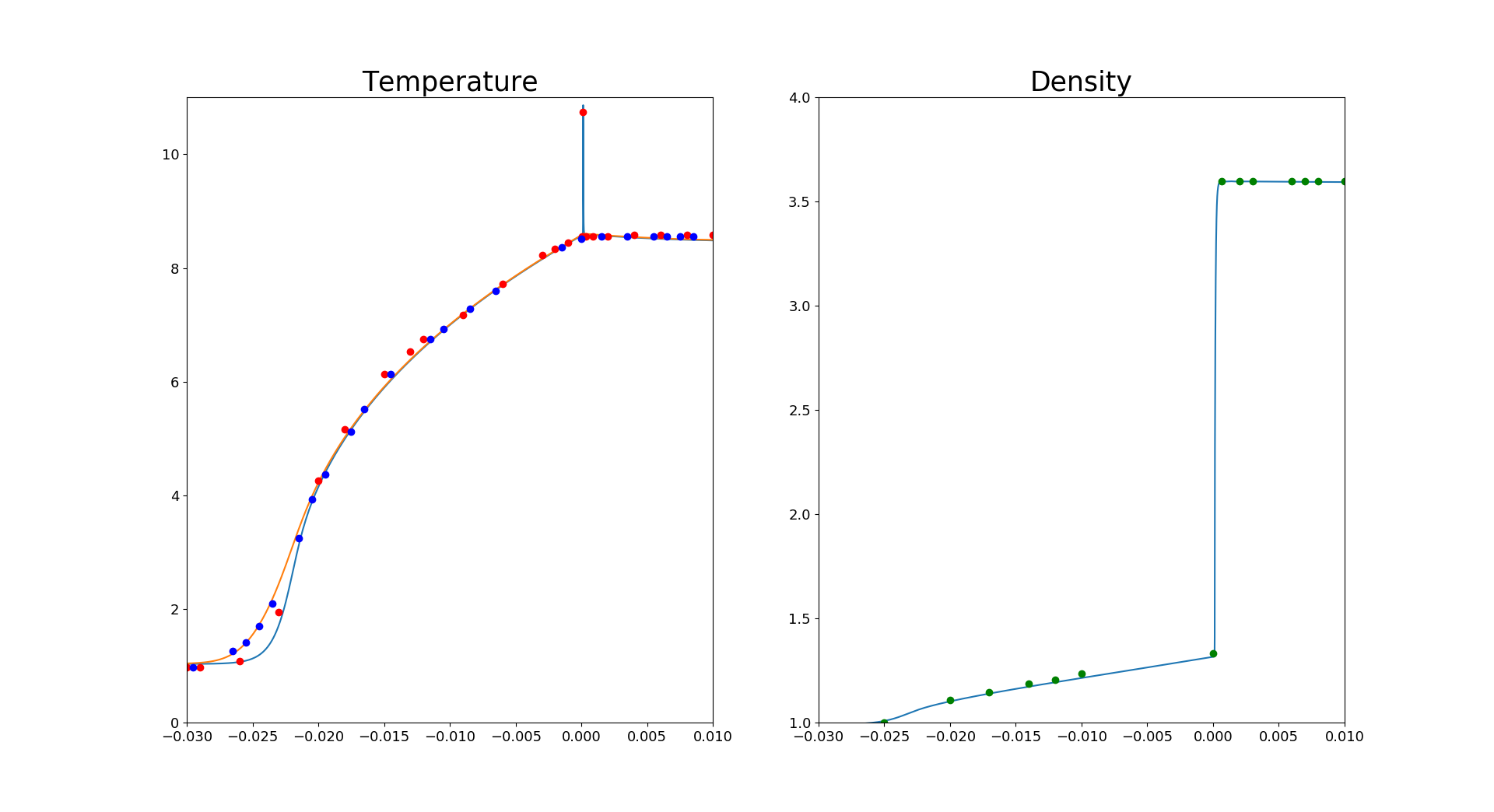}
\caption{Same as~\ref{Fig:LEmach2} but for the Mach 5 case ($\mathcal{M} =$~5).}\label{Fig:LEmach5}
\end{center}
\end{figure*}

The simulation is run for a timescale that allows the new shock structure to relax to a steady state and the final profiles, in dimensionless units,
are directly compared against the semi--analytic solutions of \citet{2008ShWav..18..129L} in Figures~\ref{Fig:LEmach2} and~\ref{Fig:LEmach5}. As
can be seen, our results are in good agreement with the semi--analytic predictions showcasing the capability of {\it RadFLAH} to handle this problem
correctly both in the subcritical and the supercritical case where the temperature spike is recovered in good precision.

\subsection{{\it Non--steady subcritical and supercritical shocks}}\label{critshocks}

Given that the treatment of radiative shocks is an important aspect of implementations like {\it RadFLAH} that are designed to study astrophysical
shocks, we opt to execute yet another similar test problem as introduced by \citet{1994ApJ...424..275E} dealing with the structure of non--steady
subcritical and supercritical shocks. This benchmark test was used to evaluate a number of previous radiation hydrodynamics implementations
\citep{2003ApJS..147..197H,2007A&A...464..429G,2014ApJ...797....4K,2015ApJS..217....9R}. In our test we adopt an initial setup nearly identical
to that presented by \citet{2014ApJ...797....4K} and compare our results against approximate analytic arguments by \citet{1984oup..book.....M}.

In this configuration the initially uniform in temperature and density fluid is compressed and a shock wave travels in the upstream direction. The hot
part of the fluid radiates thermally and as a result the radiation pre--heats the incoming (downstream) fluid. This way, a subcritical or a supercritical
shock can be formed depending on whethere there is sufficient upstream radiation flux so that the preshock and the postshock temperature
become equal. We adopt the following initial conditions: ideal fluid with $\gamma =$~5/3, $\mu =$~1.0, uniform density and temperature of
$\rho = 7.78 \times 10^{-10}$~g~cm$^{-3}$ and $T =$~10~K respectively and $\kappa_{\rm R} = \kappa_{\rm P} = 3.12 \times 10^{-10}$~cm$^{-1}$.
The domain size is $\Delta R = 7 \times 10^{10}$~cm. As with~\ref{Lowrie}, we investigate two cases: one of a subcritical shock, where the fluid
moves with $v_{\rm sh} =$~6~km~s$^{-1}$ and one of a supercritical shock with $v_{\rm sh} =$~20~km~s$^{-1}$ as in \citet{2014ApJ...797....4K}.

\begin{figure*}
\begin{center}
\includegraphics[angle=0,width=15cm,trim=0.in 0.25in 0.5in 0.15in,clip]{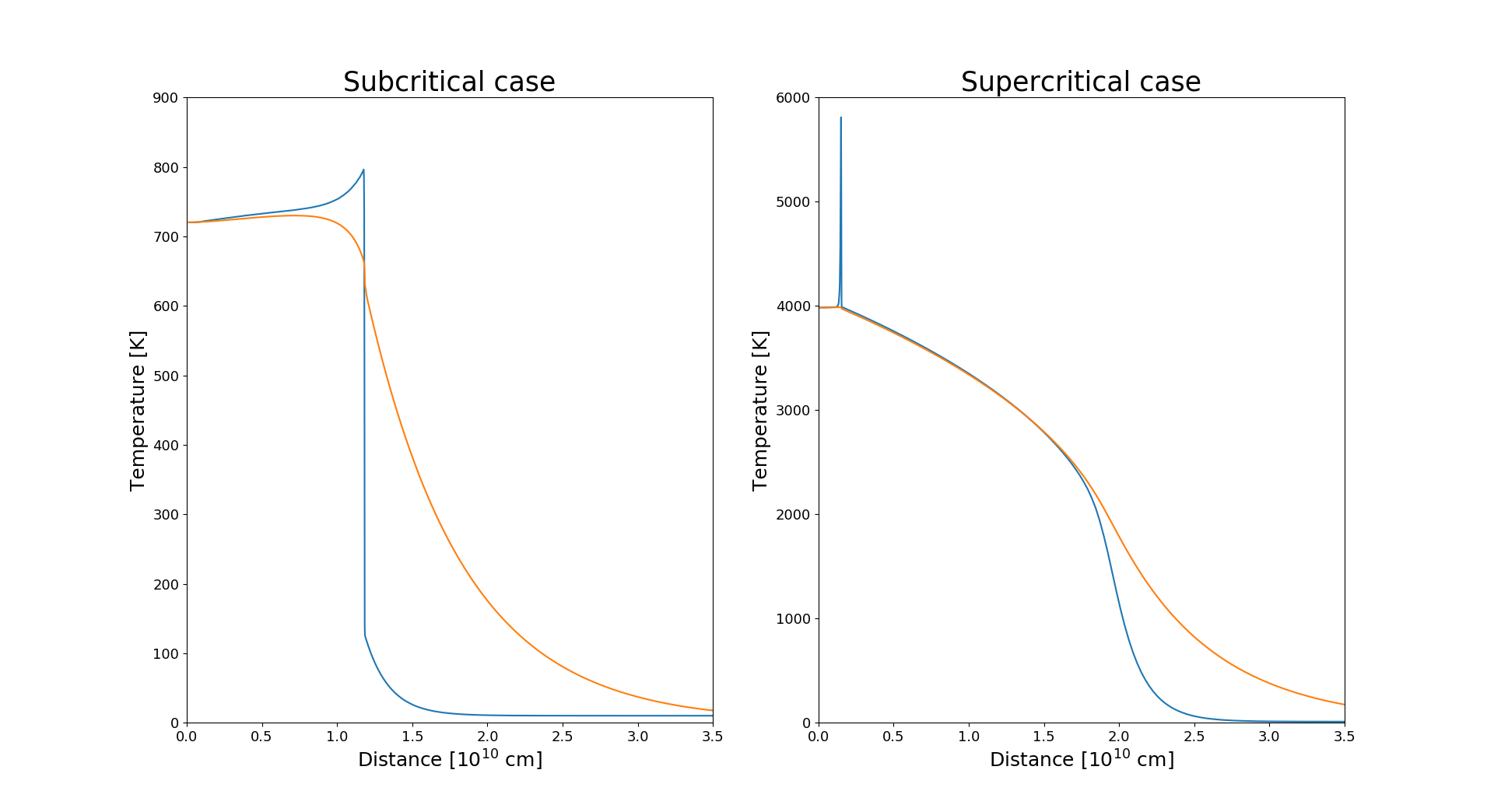}
\caption{Radiation (orange curves) and material (blue curves) temperature profiles for a non--steady
subcritical radiative shock ({\it left panel}, $v_{\rm sh} =$~6~km~s$^{-1}$, $t = 5.80 \times 10^{4}$~s)
and a non--steady critical radiative shock ({\it right panel}, $v_{\rm sh} =$~20~km~s$^{-1}$, $t = 5.08 \times 10^{3}$~s).
Our {\it RadFLAH} setup closely follows the one described in \citet{2014ApJ...797....4K} and comparison against approximate
analytical predictions is outlined in  (\ref{critshocks}).
}\label{Fig:ensman}
\end{center}
\end{figure*}

The radiation and matter temperature profiles computed in our simulation with {\it RadFLAH} are shown in Figure~\ref{Fig:ensman}. The
left corresponds to the subcritical case at $t = 5.80 \times 10^{4}$~s and the right panel to the supercritical case at $t = 5.08 \times 10^{3}$~s.
\citet{1984oup..book.....M} present approximate analytic solutions for the preshock ($T_{\rm 1}$) and the postshock ($T_{\rm 2}$)
temperature as well as the temperature spike ($T_{\rm *}$).
In the subcritical case these are given by the following expressions:

\begin{equation}
T_{\rm 1} \simeq \frac{\gamma - 1}{\rho v_{\rm sh} R} \frac{2 \sigma_{\rm B} T_{\rm 2}^{4}}{\sqrt{3}},\label{eq:preshock}
\end{equation}

\begin{equation}
T_{\rm 2} \simeq \frac{2(\gamma - 1) v_{\rm sh}^{2}}{R (\gamma + 1)^{2}}, \label{eq:postshock}
\end{equation}

\begin{equation}
T_{\rm *} \simeq T_{\rm 2} + \frac{3 - \gamma}{\gamma + 1} T_{\rm 1}, \label{eq:spike}
\end{equation}

where $\sigma_{B}$ is the Stefan--Boltzmann constant, $R = k_{\rm B}/\mu m_{\rm H}$ the ideal gas constant,
$k_{\rm B}$ the Boltzmann constant and $m_{\rm H}$ the mass of the hydrogen atom.
Using the values adopted in our test simulation, Equations~\ref{eq:preshock}, \ref{eq:postshock} and \ref{eq:spike} yield
$T_{\rm 1} \simeq$~279~K, $T_{\rm 2} \simeq$~812~K and $T_{\rm *} \simeq$~874~K accordingly. For comparison, our
simulation yields $T_{\rm 1} =$~189~K, $T_{\rm 2} =$~716~K and $T_{\rm *} =$~797~K indicating agreement within
9--32\% of the analytical estimates.
In the supercritical case the temperature spike can be approximated by:
\begin{equation}
T_{\rm *,super} \simeq \left(3 - \gamma \right) T_{\rm 2}, \label{eq:spike2}
\end{equation}
and, using the parameters adopted in our simulation corresponds to $T_{\rm *,super} \simeq$~4612~K. In contrast, our
simulation suggests $T_{\rm *,super} =$~5778~K which is within 25\% of the approximate analytical result.

The source of the discrepancies between our numerical results and the approximate analytical predictions is not due to mesh
resolution since we performed a resolution study and the same results hold in good precision. However we note that the sensitivity
to the choice of flux limiter (we use \citealt{1981ApJ...248..321L}) that controls differences in regions of
intermediate to low optical depth can account for these differences \citep{2001ApJS..135...95T}. Similar issues and
conclusions were found by \citet{2014ApJ...797....4K}.

\begin{figure*}
\begin{center}
\includegraphics[angle=0,width=15cm,trim=0.in 0.25in 0.5in 0.15in,clip]{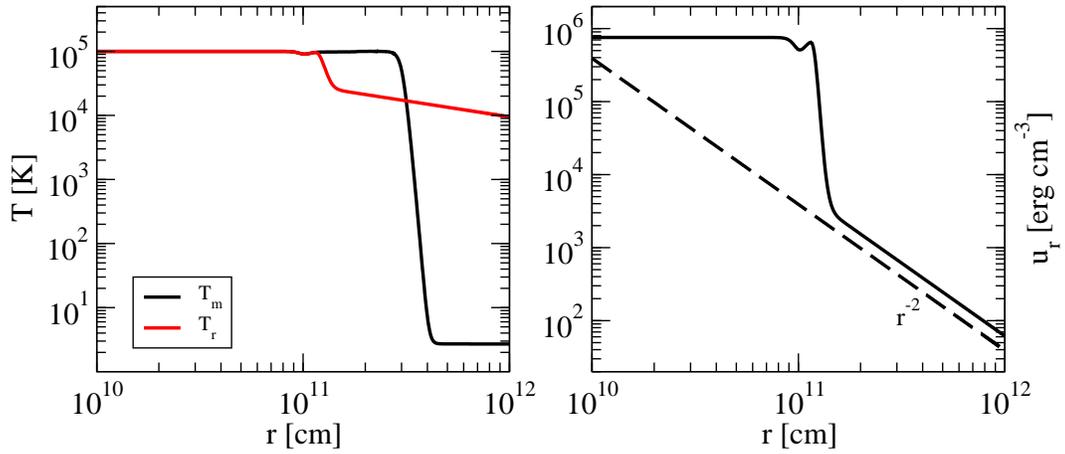}
\caption{Profiles of matter ($T_{\rm m}$; solid black curve) and radiation ($T_{\rm r}$; solid red curve) temperature ({\it left panel})
and radiation energy density ($u_{\rm r}$; solid black curve, {\it right panel}) for the radiating sphere test problem (\S~\ref{Rad_opt_thin})
at the end of the simulation ($t = 10^{6}$~s).
The dashed black curve in the right panel denotes a $u_{\rm r} \sim r^{-2}$ decline law.}\label{Fig:sigmoid}
\end{center}
\end{figure*}

\begin{figure*}
\begin{center}
\includegraphics[angle=0,width=12cm,trim=1.in 2.in 1.in 2.in,clip]{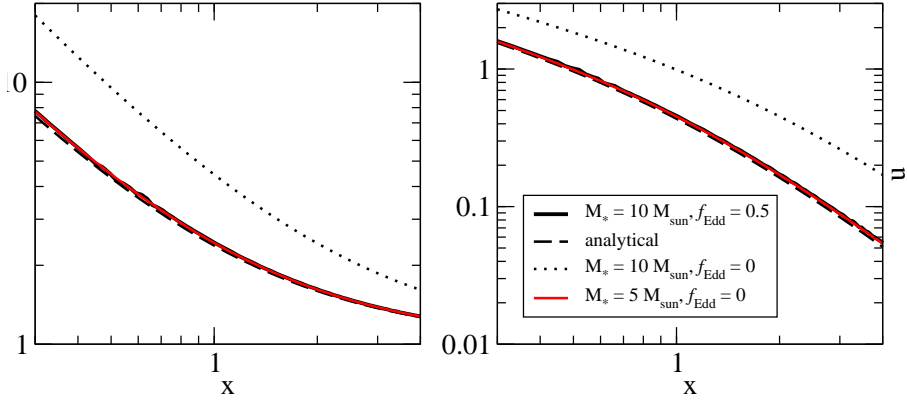}
\caption{Profiles of scaled density ({\it left panel}) and scaled radial velocity ({\it right panel}) for
the radiation-inhibited Bondi accretion test problem (\S~\ref{Bondi}) at the end of the simulation
($t = 1.538 \times 10^{7}$~s or five Bondi times). 
The solid black and red curves correspond to central source mass of 10~$M_{\odot}$ and 5~$M_{\odot}$ accordingly.
The black dashed curve shows the analytic solution and the black dotted curve the case of
accretion in the absence of radiation.}\label{Fig:bondi}
\end{center}
\end{figure*}

\begin{figure*}
\begin{center}
\includegraphics[angle=0,width=15cm,trim=0.in 0.25in 0.5in 0.15in,clip]{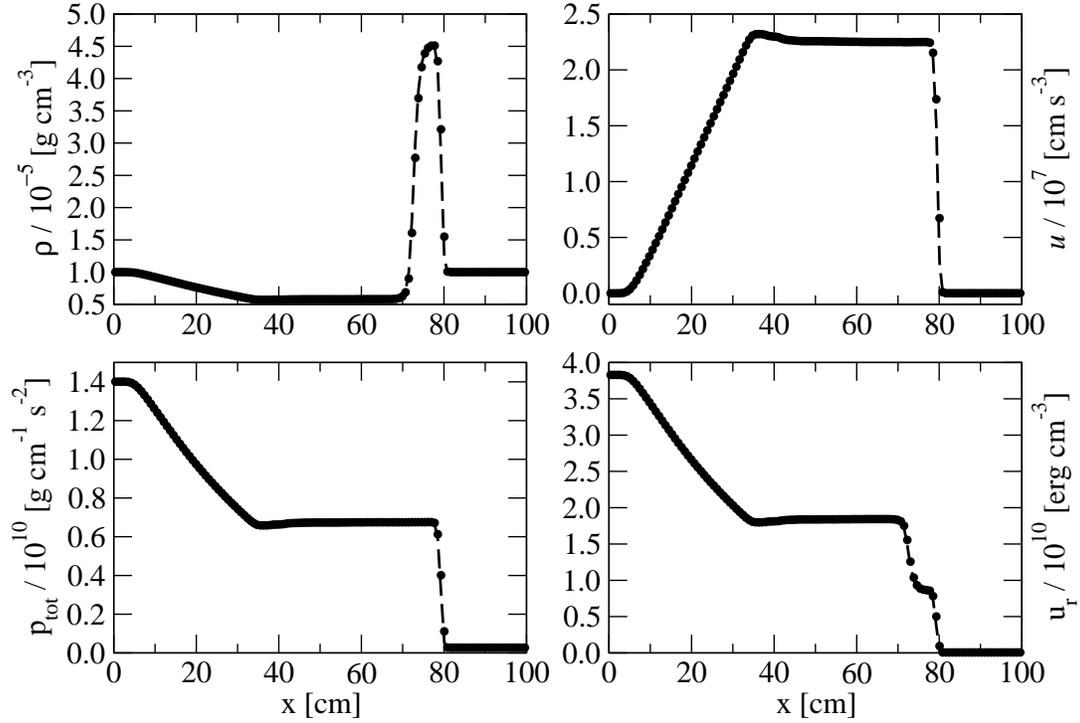}
\caption{Profiles of density ($\rho$, {\it upper left panel}), velocity ($u$,{\it upper right panel}), total pressure ($p_{\rm tot}$, {\it lower left panel})
and radiation energy density ($u_{\rm r}$, {\it lower right panel}) for the shock-tube problem in the strong coupling limit (\S~\ref{ShockTube})
at the end of the simulation ($t = 10^{-6}$~sec). Black dashed curves denote the pure hydrodynamics and filled circles the
full radiation hydrodynamics simulation.}\label{Fig:shtube}
\end{center}
\end{figure*}

\begin{figure*}
\begin{center}
\includegraphics[angle=0,width=15cm,trim=0.in 0.25in 0.5in 0.15in,clip]{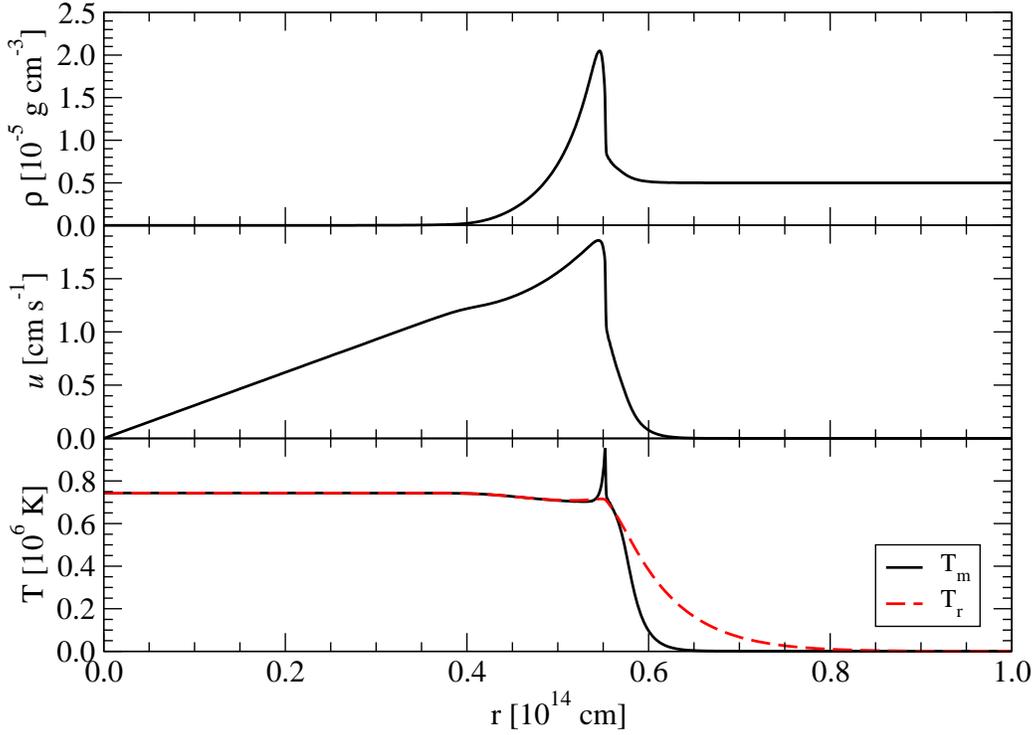}
\caption{Profiles of density ($\rho$, {\it upper panel}), velocity ($u$, {\it middle panel}) and matter ($T_{\rm m}$; black curves)
and radiation ($T_{\rm r}$; red curves) temperature ({\it lower panel}) for the radiative shock test 
problem in the weak coupling limit (\S~\ref{RadShock}) at the end of the simulation ($t = 10^{6}$~s).}\label{Fig:radsh1}
\end{center}
\end{figure*}

\begin{figure*}
\begin{center}
\includegraphics[angle=0,width=15cm,trim=0.in 0.25in 0.5in 0.15in,clip]{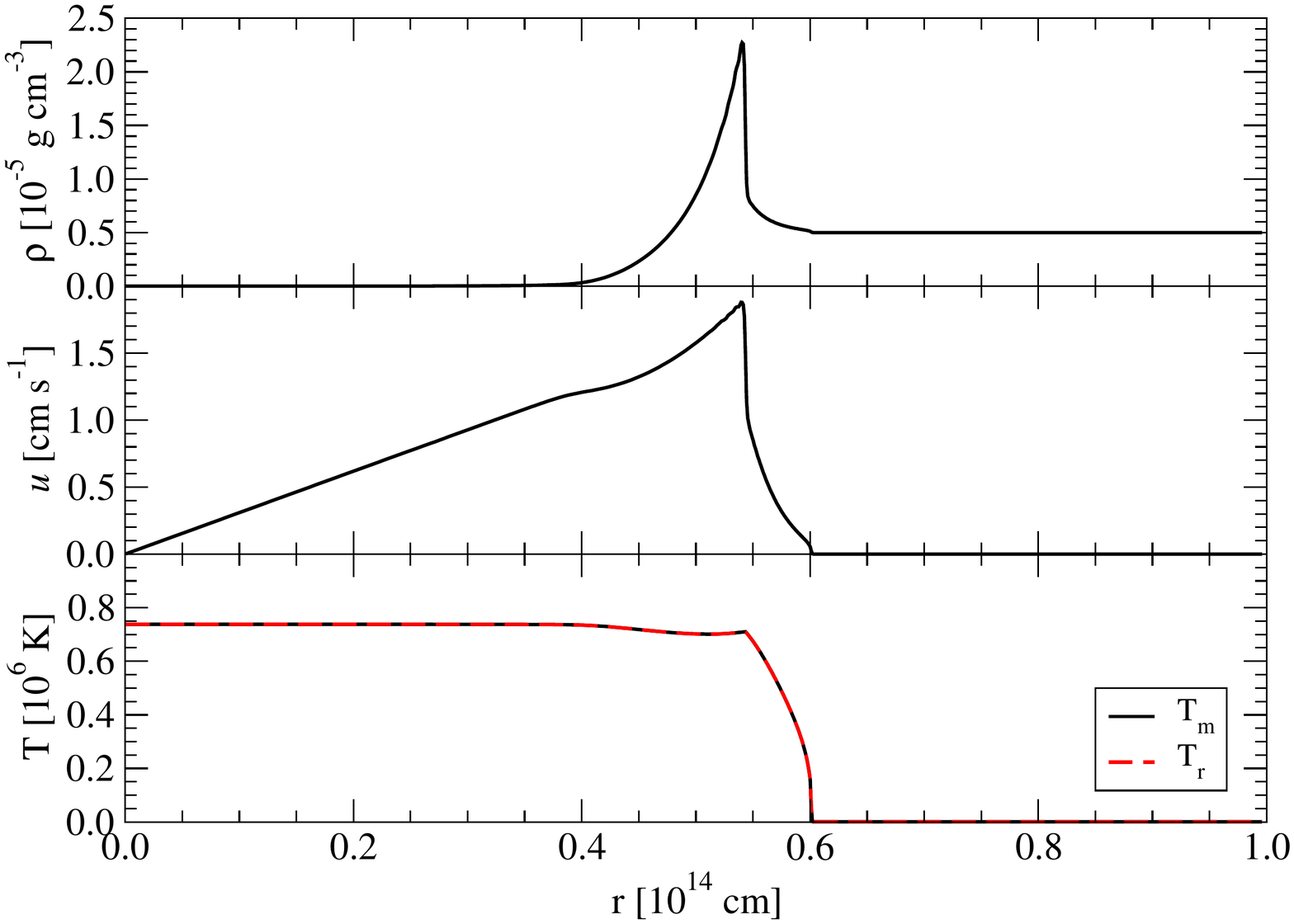}
\caption{Same as Figure~\ref{Fig:radsh1} but for the strong coupling limit (\S~\ref{RadShock}).}\label{Fig:radsh2}
\end{center}
\end{figure*}

\subsection{{\it Propagation of radiation front in the optically-thin regime}}\label{Rad_opt_thin}

In this test problem, we examine the capacity of our implementation to correctly calculate
the properties of a radiation front streaming in the optically--thin limit and its behavior at large
distances from the radiating source, tied to the outer radiation boundary conditions. 
We initialize our grid with a matter temperature and a density profile given by the sigmoid
function:

\begin{equation}
X = X_{\rm s} + \frac{X_{\rm vac} - X_{\rm s}}{1+e^{-\frac{\beta}{r_{\rm *}} (r -r_{\rm *})}},\label{eq:test1_profiles}
\end{equation} 
where $X = \rho, T_{\rm m}$ and the subscripts ``vac'' and ``s'' are used for ``vacuum'' (the outer, optically-thin region
of the domain)  and ``sphere'' (the inner, radiating sphere region) accordingly. The parameter $r_{\rm *}$
controls the radius where the profile transitions from the sphere to the vacuum and $\beta$ sets
the steepness of this transition. We select $\beta = 30$ and $r_{\rm *} = 1, 3 \times 10^{11}$~cm for
the $\rho$ and $T_{\rm m}$, accordingly. We allow the temperature profile to break at a larger radius than
the density profile in order to probe the effects of radiation matter coupling in the intermediate region.
The radiation temperature ($T_{\rm r}$) is initialized to zero throughout the domain in order to force the system to start
in an out of equillibrium state. 
We assume a fully ionized H gas that follows the $\gamma$~law equation of state (EOS) with $\gamma =$~5/3.
We also assume $\rho_{\rm s} =$~1~g~cm$^{-3}$, $\rho_{\rm vac} = 10^{-9}$~g~cm$^{-3}$, $T_{\rm s} = 10^{5}$~K and
$T_{\rm vac} =$~2.7~K. For the absorption and the transport coefficients, we set $\kappaP = 4 \times 10^{-10}$ and
$\kappaR = 4 \times 10^{-6}$~cm$^{-1}$ accordingly but use the {\tt op\_constcm2g} Opacity implementation in
{\it FLASH} that adjusts the opacity in a way that depends on the density profile given by
Equation~\ref{eq:test1_profiles} (opacity  $= \kappa/\rho$, in units of cm$^{2}$~g$^{-1}$).
For example, deep inside the sphere the transport opacity is $4 \times 10^{-6}$~cm$^{2}$~g$^{-1}$
(since $\rho_{\rm s} =$~1~g~cm$^{-3}$) while far in the vacuum it is $4 \times 10^{3}$~cm$^{2}$~g$^{-1}$
(since $\rho_{\rm vac} = 10^{-9}$~g~cm$^{-3}$). Our Rosseland and Plack mean opacity choices (\ref{T1})
imply weak coupling between radiation and matter. In addition, the material is optically--thin outside the radius of the radiating sphere. 

Figure~\ref{Fig:sigmoid} shows the final state of our simulation ($t = 10^{6}$~s). The radiation temperature has fully equillibrated with matter
temperature within the optically--thick dense sphere and the radiation energy density ($u_{\rm r}$) declines following a $r^{-2}$ law at large
distances. This is consistent with the behavior of radiative flux at large distances from a radiating source 
(the ``inverse--square law'': $u_{\rm r} = L / 4 \pi r^{2}$, where
$L$ is the intrinsic luminosity of the source and $r$ the distance from the center).

\subsection{{\it Radiation-inhibited Bondi accretion}}\label{Bondi}

To study the dynamical effects of radiation pressure on matter in the optically--thin limit we
simulate the process of radiation--inhibited Bondi accretion \citep{1952MNRAS.112..195B}.
A radiating point source of mass $M$ is assumed in the center of the domain, surrounded by
a low--density medium. Radiation from the point source
free--streams into the surrounding material exerting force on it, causing the inward spherical accretion
onto the gravitating mass to decelerate. The magnitude of the specific (per mass) radiating force on the ambient
gas is given by the following expression:

\begin{equation}
f_{\rm r} = \frac{\kappa_{R} L}{4 \pi r^{2} c},\label{eq:rad_force}
\end{equation}
where $L$ is the luminosity of the point source. 
The ratio of the radiative to the gravitational force is equal to the fraction of the Eddington luminosity with
which the central source is radiating:

\begin{equation}
f_{\rm Edd} = \frac{\kappa_{R} L}{4 \pi G M c},\label{eq:eddington}
\end{equation}
where $G$ the gravitational constant. Radiation inhibits accretion in a way that is equivalent
to the gravitational force by a non--radiating point--source with mass $\left(1-f_{\rm Edd}\right) M$. 
The time--scale for the accretion system to settle is $\simeq r_{\rm B}/c_{\rm s}$ where $r_{\rm B}$ is
the Bondi radius ($r_{\rm B} = \left(1-f_{\rm Edd}\right) GM/c_{\rm s}^{2}$) and $c_{\rm s}$ the speed
of sound in the ambient medium. Assuming an isothermal gas, analytical solutions
for the final density and velocity radial profiles can be found by solving the following system of equations \citep{1992pavi.book.....S}:

\begin{eqnarray}
x^{2} \alpha u = \xi \label{eq:bondi_one}\\
\nonumber\\
\frac{u^{2}}{2} + \ln \alpha - \frac{1}{x} = 0,\label{eq:bondi_two}
\end{eqnarray}
where $\xi = e^{1.5}/4$ is a constant specific for an isothermal gas, $x = r/r_{\rm B}$ is the dimensionless radius,
$\alpha = \rho/\rho_{\rm vac}$ the dimensionless density and $u = v/c_{\rm s}$ the dimensionless velocity.

In this test problem, we use the exact same initial setup as \citep{2007ApJ...667..626K} in order to 
compare our code with their mixed--frame implementation for radiation hydrodynamics. 
More specifically, we adopt $\rho_{\rm vac} = 10^{-18}$~g~cm$^{-3}$, $T_{\rm r, vac} = T_{\rm m,vac} = 10^{6}$~K
corresponding to $c_{\rm s} = 1.3 \times 10^{7}$~cm~s$^{-1}$. For the radiating point--source we set
$M =$~10~$M_{\odot}$ and $L = 1.6 \times 10^{5}$~$L_{\odot}$. Since we are not treating the central source as a sink particle,
in contrast with the \citet{2007ApJ...667..626K} approach, we employ the {\tt Dirichlet} option in {\it FLASH} for the inner boundary
condition for radiation, effectively fixing the radiation and matter temperature in that boundary in a way that it corresponds 
to the same $L$. We also enforce radiation--matter coupling by setting $\kappa_{\rm P} =$~0.
With this choice of parameters, $f_{\rm Edd} =$~0.5, meaning that the effects of radiation--inhibited accretion
are equivalent to pure accretion onto a non--radiating point--source with mass 5~$M_{\odot}$. 

The simulation is run for five Bondi time--scales and the results are shown in Figure~\ref{Fig:bondi}. We compare
accretion with and without radiation included for the original point source, the analytical solution and accretion without
radiation included for a point source of half mass (5~$M_{\odot}$). Our results are in very good agreement with the analytical
solution and compare well with those of \citet{2007ApJ...667..626K} (their Figure 9). 

\subsection{{\it Shock--Tube problem in the strong coupling limit}}\label{ShockTube}

To study our implementation in the limit of strong equillibrium and no diffusion we simulate
the shock--tube problem.
To compare our implementation with results from the {\it CASTRO} gray radiation
hydrodynamics framework, we use the same initial setup as the one presented in \citet{2011ApJS..196...20Z}.
We divide an 1D Cartesian grid into two distinct regions, separated in the center of the domain at 50~cm that is coincident
with a temperature discontinuity. The initial
density is uniform throughout the domain and set to $\rho(x) = 10^{-5}$~g~cm$^{-3}$. The initial velocity is zero
everywhere and the initial matter and radiation temperature are set to be equal and initialized in the following way:

\begin{equation}
T_{\rm r,m} = 1.5 \times 10^{6} \theta(50-x) + 3.0 \times 10^{5} \theta(x-50),\label{eq:t_profile_st}
\end{equation}
where $\theta(x-x^{\prime})$ is the unit step function. 
We assume the gas to be ideal ($\gamma =$~5/3) with a mean molecular weight $\mu =$~1. Due to
the large values for $\kappa_{\rm P}$, $\kappa_{\rm R}$ (\ref{T1}), matter and radiation are in strong equillibrium
and the domain is optically--thick. 

Figure~\ref{Fig:shtube} shows the final density, velocity, total (radiation plus gas) pressure and radiation energy density.
The full radiation hydrodynamics simulation (filled circles) is compared against a pure hydrodynamics simulation that
in the strong--coupling limit gives almost identical results because of the fact that
the pure hydrodynamic calculation uses an EOS which includes a radiation contribution
while the full radiation hydrodynamic calculation does not. Our results are in very good
agreement with the results presented in Figure~8 of \citep{2011ApJS..196...20Z}.

\subsection{{\it Radiative shock in the weak and strong--coupling limit}}\label{RadShock}

Given that radiative blast waves are quite common in astrophysical systems and of direct relevance
to SNe, this test problem aims to validate the capacity of our implementation to treat shocks both
in the weak and the strong radiation--matter coupling limit. 
More specifically, we evaluate our two implementations for the treatment of radiation transfer: 
the flux--limited diffusion solver presented in \S~\ref{flux_lim} and the iterative solver
for strong radiation--matter coupling (the new {\tt ExpRelax} implementation in {\it FLASH}, \S~\ref{exp_relax}).
The motivation for using  {\tt ExpRelax} in the strong coupling case is to take advantage of the reduced timesteps and
stability it offers in this regime and simultaneously test its performance as well.
To benchmark against {\it CASTRO}
we use the same simulation setups as those presented by \citet{2011ApJS..196...20Z}. Specifically,
we initialize our domain in 1D spherical coordinates and with a constant--density material, $\rho(r) = 5 \times 10^{-6}$~g~cm$^{-3}$,
at rest ($v(r) =$~0~cm~s$^{-1}$) and with a constant radiation and matter temperature set to the same value ($T_{\rm r,m} =$~1000~K).
{\bf The shock is initalized in the left (inner) part of the domain by setting both the radiation and matter temperature to
$10^{7}$~K for $r \leq 2 \times 10^{12}$~cm. This is 10,000 times higher than the temperature in the ambient material.}
We assume ideal gas ($\gamma =$~5/3) with $\mu =$~1. We select our refinement parameters in a way that corresponds to a maximum
resolution of $9.766 \times 10^{10}$~cm, intermediate between the low and high--resolution cases presented in \citep{2011ApJS..196...20Z}.

In the weak--coupling limit we take the ratio of the emission/absorption to the transport opacity to be $\kappa_{\rm P}/\kappa_{\rm R} = 10^{-6}$.
In this case, radiation is free to escape in front of the shock forming a radiative precursor and, over time, the radiation and matter temperature
depart from equillibrium. In the strong--coupling limit we take the opacity ratio to be  $\kappa_{\rm P}/\kappa_{\rm R} =$~1000. In this case, $T_{\rm r}$
and $T_{\rm m}$ remain in equillibrium throughout the simulation and the result is expected to be identical to the corresponding pure 1--T hydrodynamics case. 
Figures~\ref{Fig:radsh1} and~\ref{Fig:radsh2} show the results at the end of the simulations for the weak--coupling and the strong--coupling case
accordingly. Again, a great agreement is reproduced between the results of {\it RadFLAH} and those of  \citet{2011ApJS..196...20Z}.

\section{APPLICATION: 1D SUPERNOVA EXPLOSION}\label{SN1D}

{\bf In order to illustrate the capacity of {\it RadFLAH} to model astrophysical phenomena, we model the LCs
of SNe coming from two different progenitor stars: a red supergiant (RSG) star with an extended hydrogen envelope 
and a more compact star stripped of its hydrogen envelope (``stripped'').} The RSG model is expected to produce
a Type IIP SN LC with a long ($\sim$~100~d) plateau phase of nearly constant bolometric luminosity ($L_{\rm bol}$) followed
by the late--time decline due to the radioactive decays of $^{56}$Ni and $^{56}$Co. The ``stripped'' model on the other hand, due to the
lack of an extended hydrogen envelope and the smaller mass, will produce a {\bf fast--evolving LC with an 1-2 week long
re--brightening phase due to heating by radioactivity.}
Our model LCs will be compared against those of the
{\it SuperNova Explosion Code (SNEC)} \citep{2015ApJ...814...63M} using the same input RSG and {\bf ``stripped''} SN profiles.  

\subsection{{\it Heating due to radioactive decay of  $^{56}$Ni and SN ejecta opacity}}\label{rad_decay}

\begin{figure*}
\begin{center}
\includegraphics[angle=0,width=15cm,trim=0.in 0.25in 0.5in 0.15in,clip]{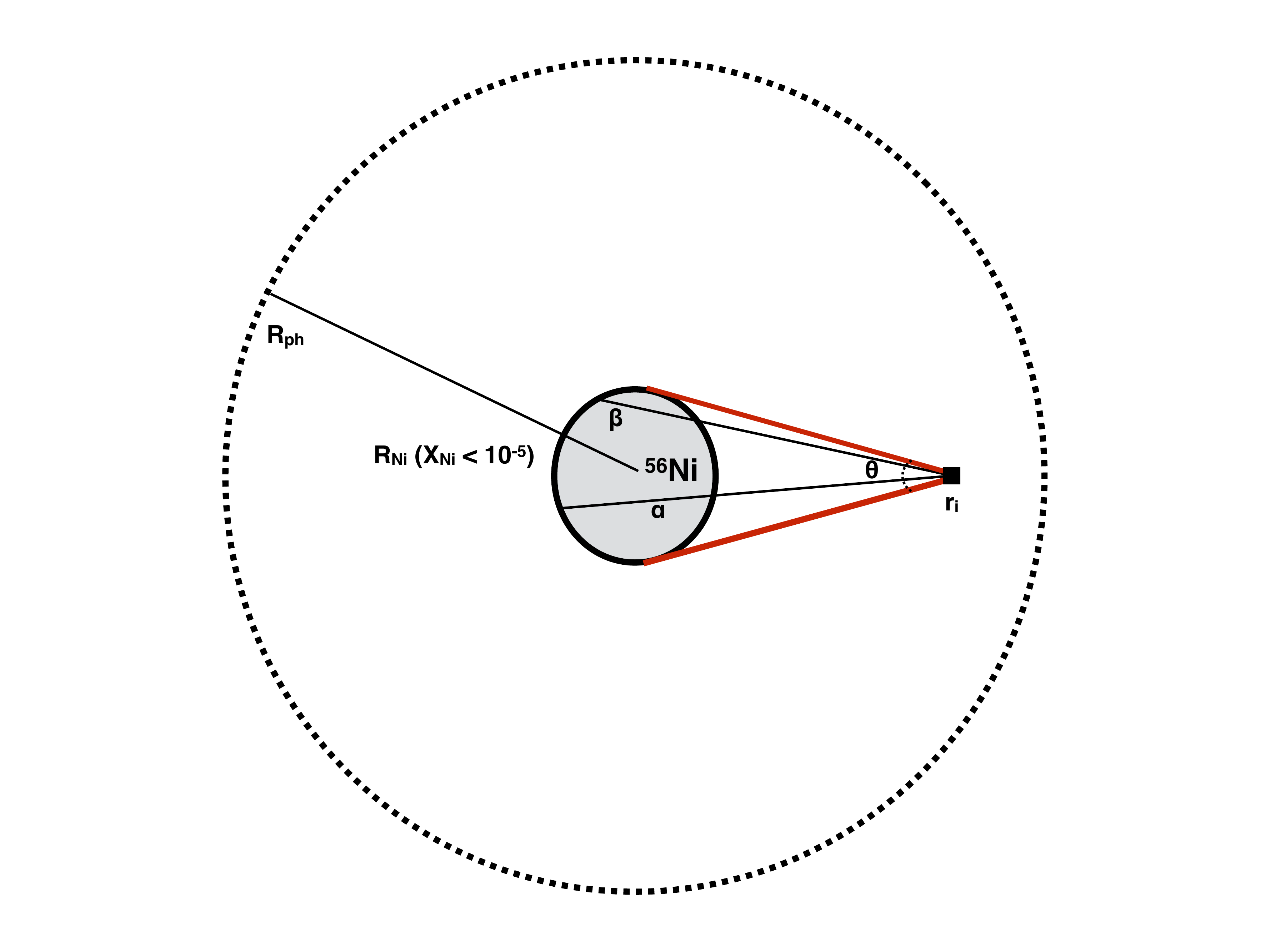}
\caption{Illustration of the method used to calculate updated specific internal energy in the star due to heating by the $^{56}$Ni and $^{56}$Co radioactive decays.
$R_{\rm ph}$ and $R_{\rm Ni}$ refer to the radii of the photosphere and the $^{56}$Ni sphere respectively defined by the location where $X_{\rm Ni} < 10^{-5}$. 
The zone in location $r_{\rm 1}$ in the SN ejecta is heated by the decay of radioactive material spanning an angle $\theta$. 
We consider an angular resolution ($\theta/N_{\rm angular}$), where $N_{\rm angular}$ the number of rays extending from the $^{56}$Ni sphere to $r_{\rm 1}$. 
For each path (example paths $\alpha$ and $\beta$ are shown, we also consider a ``radial resolution'', $N_{\rm radial}$, along the path to sum 
contributions due to heating from all regions of the $^{56}$Ni sphere.}\label{Fig:cartoon}
\end{center}
\end{figure*}

A new {\tt Heat} physics unit was implemented in {\it FLASH} to treat the heating of the SN ejecta due to
$\gamma$--rays produced by the radioactive decays of $^{56}$Ni and $^{56}$Co. The method used to re--calculate
the specific internal energy added in each zone is entirely based on \citet{1995ApJ...446..766S} 
and it is the same technique incorporated in {\it SNEC} and described in the code's users guide online 
\footnote[1]{https://stellarcollapse.org/SNEC}. 

This method involves solving the radiation transfer equation in the gray approximation assuming
$\gamma$--ray opacity, $\kappa_{\rm \gamma} = 0.06 Y_{\rm e}$~cm$^{2}$~g$^{-1}$, where $Y_{\rm e}$ is the electron
fraction. The algorithm loops through all radial zones and calculates the integrated intensity of radiation coming
from paths originating from a central spherical region where $^{56}$Ni is concentrated (Figure~\ref{Fig:cartoon}). To determine the radius
of the $^{56}$Ni sphere we set a threshold on the $^{56}$Ni mass fraction of $10^{-5}$. We then define a radial ($N_{\rm radial}$)
resolution along each path and an angular ($N_{\rm angular}$) resolution that determines the number of paths originating
from the $^{56}$Ni sphere that contribute to the heating of each zone. For the models discussed later we
use $N_{\rm radial} = N_{\rm angular} =$~100. Finally, the internal energy of each zone is updated accordingly by adding that
extra heating source term. To preserve a fast running--time, we only add the radioactive decay heating periodically,
every one day (86,400~s) throughout the run. 

Given our objective to model radiation diffusion through SN ejecta, a new {\it FLASH} {\tt Opacity} was developed that takes advantage of the 
{\it Lawrence Livermore National Laboratory (LANL) OPAL} opacity database \citep{1996ApJ...464..943I}. 
We specifically used opacity tables in two temperature regimes: the low ($\log T < 4.5$; \citealt{2005ApJ...623..585F}) and the high--temperature
($\log T > 4.5$; \citealt{1998SSRv...85..161G}) regime based on solar metal abundances. 
{\bf We directly linked the {\it OPAL} tables from the stellar evolution {\it MESA} code opacity database in order to take advantage of
the consistent and succinct formatting in these files. This way, all values for the Rosseland mean opacity directly correspond to
the {\it OPAL} values for each zone in the initialization of the supernova runs. For the Planck mean opacity, on the other hand, we
adopted a fiducial constant value by assuming Thompson scattering as the main source of opacity. As such, for the (H--rich)
``RSG'' run we have used $\kappa_{\rm P} =$~0.4~cm$^{2}$~g$^{-1}$ and for the (H--poor) ``stripped'' run  
$\kappa_{\rm P} =$~0.2~cm$^{2}$~g$^{-1}$.}
In order to be provided with a robust comparison against the results of {\it SNEC}, 
we had to impose their adopted opacity floor given by:
\begin{equation}
\kappa_{\rm floor} (r) = \frac{0.24 Z_{\rm env} - 0.01 - 0.23 Z(r)}{Z_{\rm env} - 1},
\end{equation}
where $Z_{\rm env}$ is the metallicity of the stellar envelope and $Z(r)$ the metallicity as a function of radius.

\subsection{{\it Input SN ejecta profiles}}\label{sn_profs}

\begin{figure*}
\begin{center}
\includegraphics[angle=0,width=18cm,trim=0.in 0.25in 0.5in 0.15in,clip]{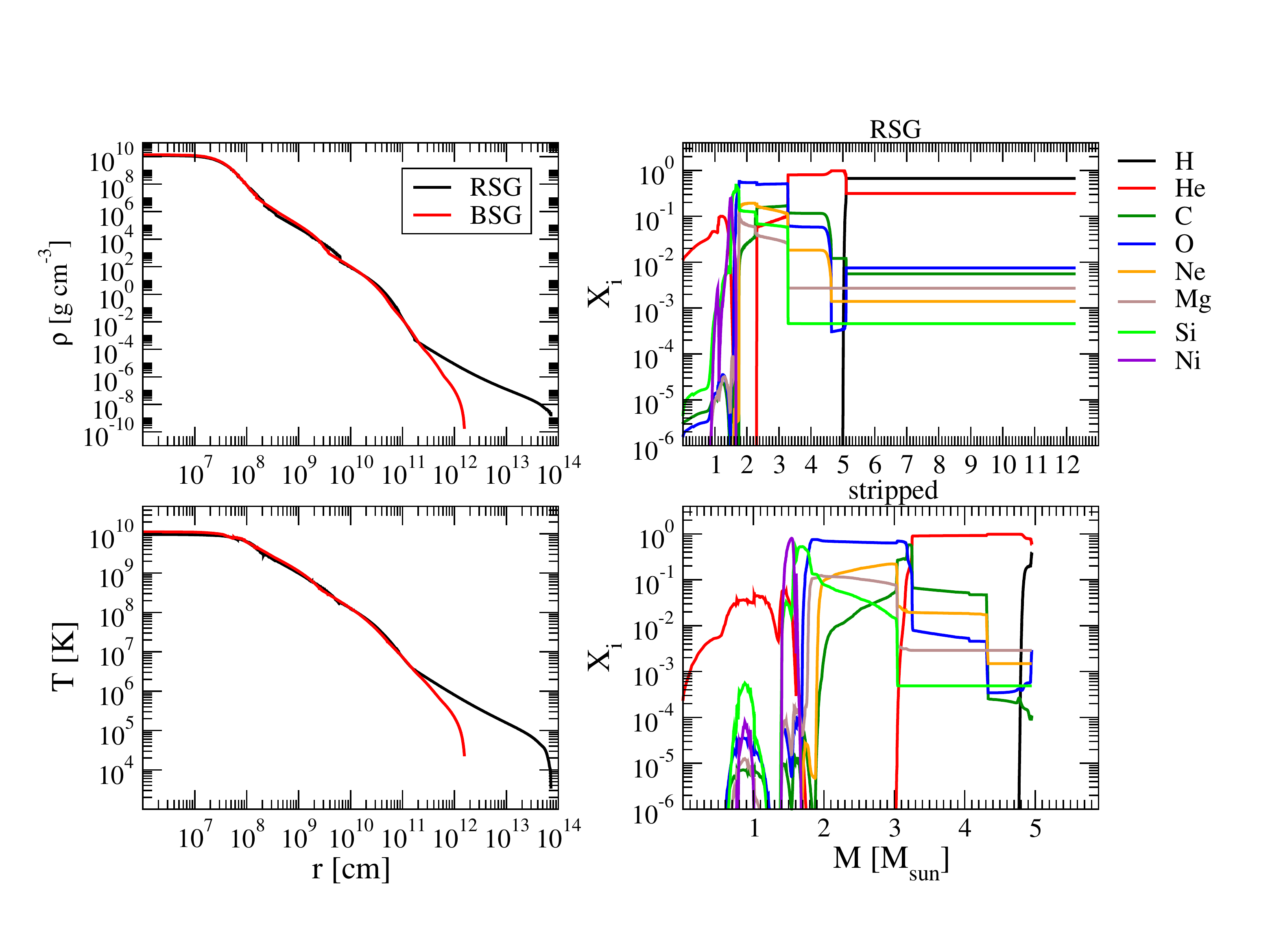}
\caption{The inital SN profiles used for the calculation of gray LCs with {\it FLASH RadFLAH}. Density ({\it upper left panel}) and temperature ({\it lower left panel}) for
the RSG (black curves) and {\bf ``stripped''} (red curves) models. Composition profiles for the RSG ({\it upper right panel}) and the {\bf ``stripped''} ({\it lower right panel}) models where
no boxcar mixing is applied (\ref{sn_profs}).}
\label{Fig:SNprofiles}
\end{center}
\end{figure*}

Figure~\ref{Fig:SNprofiles} shows the initial structural properties ($\rho$, $T$ and composition) of the basic RSG and {\bf ``stripped} models
used taken from the available profiles within the {\it SNEC} source tree ({\tt 15Msol\_RSG} and {\tt stripped\_star} therein). 
In {\it SNEC} it is emphasized that these models were evolved to the pre--SN stage using the {\it MESA} code. 
The RSG model represents a red supergiant star that was 15~$M_{\odot}$
at Zero Age Main Sequence (ZAMS) while the {\bf ``stripped model} a compact blue star from a 15~$M_{\odot}$ ZAMS model where the convective
envelope was stripped during the evolution \citep{2014ApJ...792L..11P}. Considering mass--loss during the evolution, the final,
pre--explosion models had total masses of 12.2~$M_{\odot}$ (RSG) and 4.9~$M_{\odot}$ ({\bf ``stripped''}).

{\it SNEC} provides the user with the option to set a total $^{56}$Ni mass as an input and the option to apply one--dimensional 
parameterized mixing due to the Rayleigh--Taylor and Richtmyer--Meshkov instabilities the SN ejecta using the boxcar smoothing method 
\citep{2009ApJ...703.2205K}. In order to investigate these effects we run three {\it SNEC} models for each progenitor: one
with $M_{\rm Ni} =$~0.05~$M{\odot}$ using the original SN ejecta profiles,
one with $M_{\rm Ni} =$~0.05~$M{\odot}$ but with boxcar smoothing applied,
and one with no $^{56}$Ni radioactive decay contributions
for a total of six {\it SNEC} models.
For all three RSG' models and the {\bf ``stripped''} model with boxcar mixing applied run in {\it SNEC}, we extract density, temperature and velocity profiles at a time prior to SN shock
break--out and when the shock front is a few tenths of a solar mass within the photosphere (taken to be at optical depth of 2/3). 
Also, since {\it SNEC} does not use nuclear
reaction networks and no nucleosynthesis is performed after the explosion, the initial input model abundance profiles are assumed fixed except
for the models for which modifications were applied using boxcar averaging.
All {\it SNEC} pre--SN break--out profiles are then mapped into the 1D Adaptive Mesh Refinement ({\it AMR})
grid of {\it FLASH} and their evolution is modeled using the {\it RadFLAH} implementation yielding the computation of gray LCs.
{\bf We note that in the latest release of RadFLAH we have also included the capability for the user to initiate a ``thermal bomb''--driven
explosion in the inner regions of the initial SN profile without having to do that step within another code like {\it SNEC}.}

For the {\it FLASH} simulations we used a simulation box of length {\bf $4 \times 10^{16}$~cm}, large enough to follow the expansion of the SN ejecta
for a few hundred days. For this reason we had to inlcude a low--mass circumstellar wind with density scaling as $r^{-2}$ outside the star. The
temperature of the wind was kept constant at 100~K and the composition was taken to be the same as that of the outer zone of the stellar model.
{\bf The wind was constructed by assuming a mass--loss rate of $10^{-5}$~$M_{\odot}$~yr$^{-1}$ and a wind velocity of 250~km~s$^{-1}$.
The density of the wind followed an $\sim r^{-2}$ profile consistent with the observed properties of RSG--type winds 
(see Figure 3 in \citet{2014ARA&A..52..487S}).
The presence of wind material around the SN progenitor makes the effects of the interaction between the SN ejecta and that wind inevitable, 
yet minimized in our runs given the relatively low wind density and total mass. 
For a more thorough review on the effects of pre--SN winds for high mass--loss
rates ($> 10^{-4}$~$M_{\odot}$~yr$^{-1}$) on the LCs of SNe the reader is encouraged to review \citep{2014MNRAS.439.2917M}.
To calculate the bolometric gray LCs in {\it RadFLAH}, a photosphere--locating algorithm
was employed that tracks the location of the optical depth $\tau =$~2/3 surface over time and uses the local conditions there to estimate the 
emergent luminosity.}

\subsection{{\it SN lightcurves with RadFLAH}}\label{SNSBO}

\begin{figure*}
\begin{center}
\includegraphics[angle=0,width=18cm,trim=0.in 0.25in 0.5in 0.15in,clip]{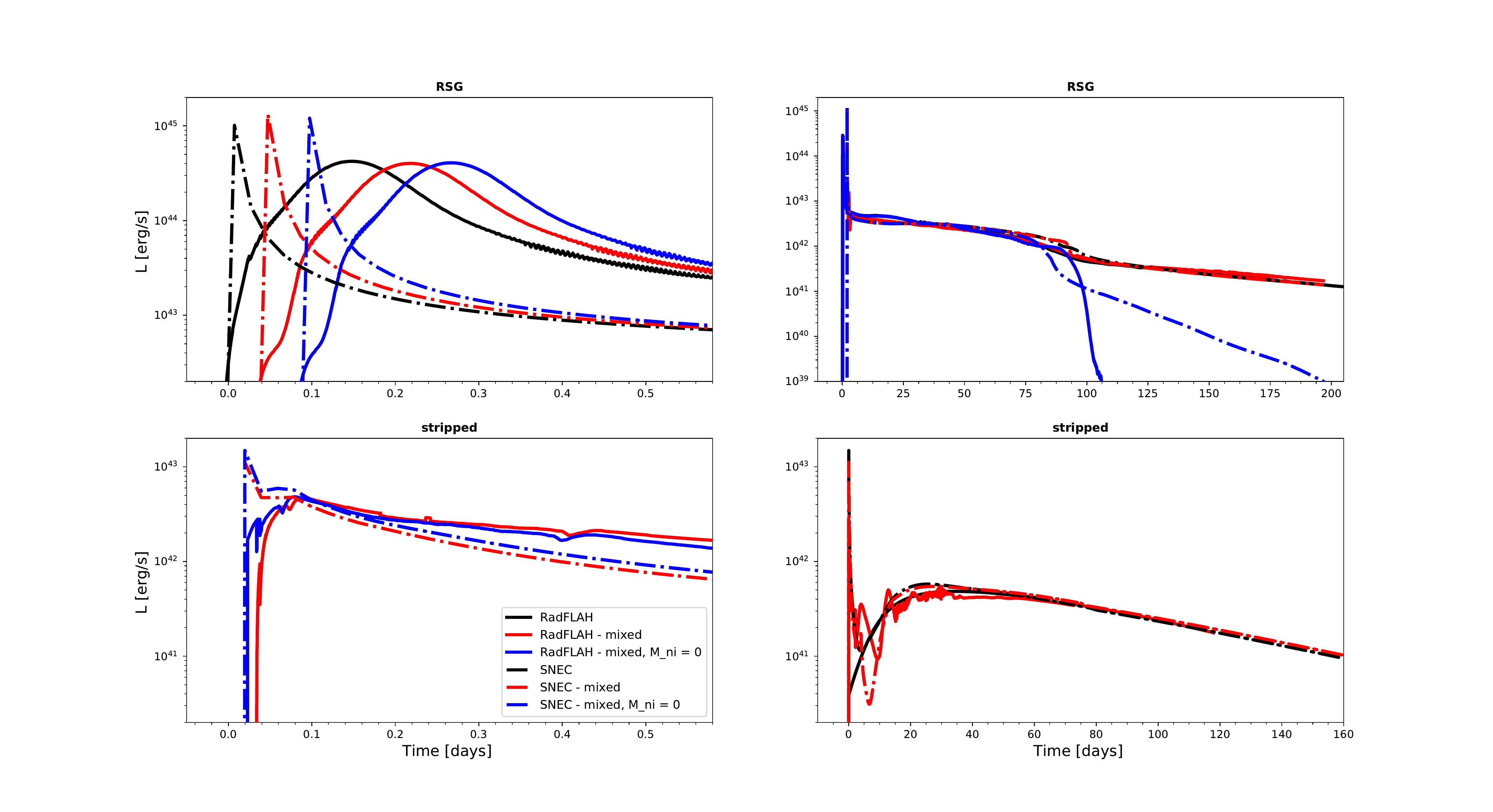}
\caption{Comparison between {\it FLASH RadFLAH} (solid curves) and {\it SNEC} (dashed curves) SN LCs. 
The upper panels show the results for the ``RSG'' models and the lower panels those for those {\bf ``stripped''} model. 
The left panels show a 0.8~day zoom--in the early shock--breakout LCs while the right panels the full LC. The agreement between the 
two codes is reasonably good given differences in the numerical treatment of radiation diffusion and microphysics.}
\label{Fig:SNLCs}
\end{center}
\end{figure*}

Figure~\ref{Fig:SNLCs} shows comparisons between the {\it SNEC} and {\it FLASH RadFLAH} LCs for the RSG (upper panels) and {\bf ``stripped''} (lower panels) models.
The left panels are a zoom--in to the early shock break--out and ``fireball'' expansion phase while the right panels show the total LC evolution, including
re--heating of the SN ejecta due to the radioactive decay of $^{56}$Ni. The comparison between the shock--breakout LCs indicates that the {\it FLASH RadFLAH}
models exhibit a less luminous yet longer--lasting break--out phase for both the RSG and {\bf ``stripped''} model, 
although the total radiated energy is about the same. These differences are attributed
to two factors. First and foremost, the two--temperature (2T) treatment where we allow the material and radiation temperature to de--couple in {\it RadFLAH} while
there is just one combined temperature used in {\it SNEC}. During shock break--out in SNe, 2T effects are strong in the weak coupling limit (see also
\ref{RadShock}). This includes the effect of a radiative precursor leaking ahead of the shock and heating the surrounding medium thus driving the radiation
temperature at the photosphere to lower values. Secondly, in contrast with the {\it SNEC} setup we include a low--density wind around the star that can
also influence the properties of shock brek--out emission.

The later, re--brightening phases due to the deposition of gamma--rays to the SN ejecta by $^{56}$Co decay are in good agreement with the {\it SNEC} results for
both models. The $\sim$~100~d plateau phase for the RSG models is reproduced at a luminosity of $\sim$~$3 \times 10^{42}$~erg~s$^{-1}$ that is typical for
Type IIP SN LCs. Also, the late--time ($>$~100~d) radioactive decay tail that has a characteristic constant decline rate for $^{56}$Co is reproduced and is consistent
with the {\it SNEC} results. For the RSG model with $M_{\rm Ni} =$~0 there are considerable differences between the {\it SNEC} and {\it FLASH RadFLAH} results at
late times after the plateau, with the {\it FLASH RadFLAH} models exhibiting a much faster decline in luminosity. The {\it FLASH RadFLAH} result is more in line with
the predictions of analytical models for Type IIP LCs like that of \citet{1989ApJ...340..396A}, 
given that the effective opacity drops to zero after the end of the hydrogen recombination phase and luminosity should quickly decline during
the nebular phases. Similar ``tail--less'' Type IIP SN LC models in the context of pulsational pair--instability explosions from massive progenitor stars
were computed by \citet{2017ApJ...836..244W} featuring rapid decline rates once the hydrogen recombination front recedes inwards. Another source of this
discrepancy is the post--plateau opacities adopted in the {\it SNEC} code attempting to take into account effects due to dust formation in the SN ejecta at
late times and low--temperature conditions \citep{2008IAUS..252....1F}.

The {\bf ``stripped'' LC models are also in good agreement between the two codes and are characterized by a faster LC evolution attributable to the smaller
initial radius and envelope mass for these progenitors.} The same effect of a more smeared--out
shock break--out LC is observed here as was the case for the ``RSG'' model but the later evolution and the $^{56}$Ni decay tail are in great agreement
between {\it SNEC} and {\it FLASH RadFLAH}. 

Given the many differences in the treatment of radiation diffusion between the two codes, the initial setup requiring the presence of a circumstellar wind
in {\it FLASH} and discrepancies in the overall numerical implementation, the agreement between the two codes is intriguing and illustrates the capacity of the
new {\it RadFLAH} implementation to provide basic 2T modeling for explosive astrophysical flows including SNe and interaction of SN ejecta with circumstellar
matter (CSM).

\section{DISCUSSION}\label{Disc}

The multi--physics, multi--dimensional AMR code {\it FLASH} has been used for studies of the hydrodynamics of astrophysical systems extensively
in the past \citep{2005Ap&SS.298...25C,2013ApJ...776..129C,2013ApJ...778L...7C,2014ApJ...795...92C,2014ApJ...797....4K,2015ApJ...799....5C,
2015ApJ...808L..21C,2016ApJ...822...61C,2016ApJ...823...28K}. Although a three--temperature (electron, ion and radiation temperature) 
radiation diffusion scheme was already present in {\it FLASH},
it was tailored for the treatment of high energy density and laser physics problems and direct application for physical regimes that are appropriate for astrophysical
objects like supernovae was not feasible. 

For this reason, we extended the hydrodynamics capabilities of the unsplit hydrodynamics solver available in {\it FLASH} and implemented the new Radiation
Flux--limiter Aware Hydrodynamics ({\it RadFLAH}) framework able to treat astrophysical problems by evolving the radiation and matter separately in a 
two--temperature approach and in the gray approximation using the Levermore--Pomraning approximation for the flux limiter. 

To be able to utilize our method for astrophysical applications, we implemented an extension of the existing ``Helmholtz''  equation of state in {\it FLASH} 
to lower temperature and density regimes characteristic of stellar photospheres and circumstellar environments. We also introduced a new
opacity unit linking the {\it OPAL} opacity database to obtain transport opacity values as a function of local temperature, density and composition. Finally,
we introduced a commonly--used method to treat the deposition of gamma--rays to the SN ejecta due to the $^{56}$Co and $^{56}$Ni radioactive decay 
heating as necessary in order to calculate complete SN LCs to late times after the explosion. 

We compared {\it FLASH RadFLAH} to flux--limited diffusion implementation used in other codes like {\it CASTRO} \citep{2011ApJS..196...20Z} and the \citet{2007ApJ...667..626K} 
code as well as analytical solutions by running standard radiation hydrodynamics and radiation diffusion test problems identical to some of those presented in their methods 
papers and found very consistent results. 
Finally, we performed a direct code--to--code comparison with the Supernova Explosion Code ({\it SNEC} \citet{2015ApJ...814...63M}) in order to assess our computed SN
LCs for two modes: a red supergiant progenitor with an extended hydrogen envelope and a more compact blue supergiant progenitor that experienced
strong mass--loss during its evolution, originally performed with the {\it MESA} stellar evolution code. Given differences in the numerical treatment of hydrodynamics
(two--temperture in {\it RadFLAH} versus one--temperature in {\it SNEC}) and radiation transfer as well as initial setup (in {\it FLASH} we had to use a large 
simulation box and provide data for a low--density circumstellar wind around the progenitor star models), {\it RadFLAH} LCs were consistent with 
those computed by {\it SNEC} for the same inital SN profiles. 
More specifically, we were able to reproduce the characteristics of the main (post break--out) and late--time (radioactive decay ``tail'') phase for both models very well. 
The differences due to our two--temperature treatment and the existence of a low--density wind around the progenitor causing some SN ejecta--circumstellar matter
interaction effects, are more prevalent during the early bright shock--break out phase of the LCs. More specifically, we computed shock--break out LCs that reach lower peak
luminosities and last longer than the ones found by {\it SNEC}, yet the total radiated energy throughout this early burst remained consistent. 

\subsection{Applicability of RadFLAH approach}
The {\it RadFLAH} method is applicable to a variety of astrophysical radiation hydrodynamics problems beyond simple SN LC computations like, for example, 
studies of SN ejecta--CSM interaction. In a future release we plan to expand the {\it RadFLAH} capabilities to treat problems in two and 
three dimensions and for different geometries as well as to incorporate a multi--group treatment for radiation diffusion allowing the user to compute band--specific SN LCs.
Given the open access to the public release of the {\it FLASH} code and its popularity amonst numerical astrophysicists,
we hope that this new, open framework finds good use in the community. 

Based on the associated approximations and assumptions, we expect our method to be particularly useful in regimes that are either
close to diffusive or close to free--streaming. 
The accuracy and stability of the method under conditions of dynamical diffusion (for example, when
$v/c << 1$ does not apply) has not been examined and should not be assumed.
We expect the method to give good solutions in diffusion--dominated and free--streaming regions
of a simulation domain; and to sensibly connect such different regions if they exist.
We do not expect the solution to particularly good in regions that cannot be viewed
as close to either (statically) diffusive or free--streaming radiation.

Stability of simulations is not always given, in particular due to the time--lagged handling
of some quantities in the equations (in particular the flux limiter $\lambda$).
This is subject to further research.

\acknowledgments

We would like to thank Daan van Rossum, Mikhail Klassen, Sean M. Couch, Donald Q. Lamb, 
Carlo Graziani, Petros Tzeferacos, Michael Zingale and the {\it SNEC} development team 
for useful discussions and comments. We would also like to thank our anonymous referee
for offering important constructive criticism and suggestions that significantly improved
our manuscript.
This work was supported in part at the University of Chicago
by the U.S. Department of Energy (DOE) under contract B523820
to the NNSA  ASC/Alliances Center for Astrophysical Thermonuclear
Flashes; the U.S. DOE NNSA ASC through the Argonne Institute for Computing in Science under field
work proposal 57789; and by the National Science Foundation under grant 
AST--0909132. The software used in this work was in part developed
by the DOE NNSA--ASC OASCR Flash Center at the University of Chicago.


\bibliography{references}

\section*{APPENDIX}\label{AppendixA}

\newcommand{\m}{{\mathrm m}}
\renewcommand{\r}{{\mathrm r}}
\newcommand{\btwom}{\alpha_{\mathrm m}}
\newcommand{\btwor}{\alpha_{\mathrm r}}
\newcommand{\btwoc}{\alpha_c}
\newcommand{\btwomr}{\alpha_{\rm m,r}}
\newcommand{\bzerom}{\zeta_{\mathrm m}}
\newcommand{\bzeror}{\zeta_{\mathrm r}}
\newcommand{\bzeroc}{\zeta_c}
\newcommand{\bzeromr}{\zeta_{\rm m,r}}
\newcommand{\bonem}{\beta_{\mathrm m}}
\newcommand{\boner}{\beta_{\mathrm r}}
\newcommand{\bonec}{\beta_c}
\newcommand{\bonemr}{\beta_{\rm m,r}}
\newcommand{\Utilde}{{\mathitbf{\widetilde U}}}
\newcommand{\vvtilde}{{\mathitbf{\tilde \vv}}}

To discuss our implementation in more detail, in a 2T(M+R) formulation, we write our state in (mostly) conservative form as
introduced above,
\begin{equation}
{\mathitbf{U}} = \begin{pmatrix}
\rho\cr
\rho\vv\cr
\Etot\cr
\rho \emat\cr
\Erad\cr
X_1\rho\cr
\vdots\cr
X_n\rho
\end{pmatrix}
\end{equation}
and our evolution equations as
\begin{equation}
\frac{\partial}{\partial t}{\mathitbf{U}} = f_{hyperbolic}+f_{fixup1}+f_{fixup2}+f_{Lorentz}+f_{transp} \,.
\end{equation}

To allow for different choices for the implementation of some terms, and
allow for parametric control of these for the purpose of experimentation,
we introduce numerical parameters
$
\btwom,\btwor,
\bonem ,\boner 
\in [0,1]$.
These control, separately for both matter and radiation components of energy, whether 
(and, if we allow them to have non-integer values, to what degree):
\begin{itemize}
\item pressure terms are included in the conservative fluxes ($\btwomr$),
\item work terms are implemented explicitly ($\bonemr$),
\end{itemize}
and we require
$$\btwoc + \bonec \le 1   \quad\mbox{for $c\in\{{\mathrm m},{\mathrm r}\}$}.$$
In case we want the dominant changes of $\Emat,\Erad$ that go beyond simple advection
to be completely represented by explicit
terms in $f_{hyperbolic}$ and $f_{fixup1}$, we have to set $\btwoc + \bonec=1$.
If, on the other hand, we want those changes to be handled by the $f_{fixup2}$ term,
we set $\btwoc = \bonec=0$.
For the tests presented in this paper, we have typically chosen either the latter or 
$\beta_{r} =$~0, $\beta_{m} =$~1, $\alpha_{m} =$~0, $\alpha_{r} =$~1.

Then
\begin{equation}
f_{hyperbolic} =
\begin{pmatrix}
-\nabla\cdot(\rho\vv)\cr
-\nabla\cdot(\rho\vv\vv) - \nabla \pmat - \lambda \nabla \Erad \cr
-\nabla\cdot \left[
(\Etot+\ptot + \pcorr)\vv
\right]
\cr
-\nabla\cdot \left(
\rho \emat \vv
+\btwom  \pmat \vv
\right) \cr
-\nabla\cdot \left[
(1+\btwor \lambda)\Erad\vv+\pcorr\vv
\right] \cr
-\nabla\cdot(\rho X_1 \vv)\cr
\vdots\cr
-\nabla\cdot(\rho X_n \vv)\cr
\end{pmatrix},
\end{equation}
\begin{equation}
f_{fixup1} =
\begin{pmatrix}
0\cr
\mathbf{0}\cr
0 
\cr
-\bonem \pmat\,\nabla\cdot \vv 
+\btwom  \vv\cdot\nabla\pmat
\cr
- \boner \lambda\Erad \, \nabla\cdot \vv
-
( 1 - \btwor )
 \left(\Erad\vv\cdot
\nabla\lambda
\right) 
+\,\btwor 
\lambda\, \vv\cdot
\nabla\Erad
 \cr
0\cr
\vdots\cr
0
\end{pmatrix},
\end{equation}
\begin{equation}
f_{fixup2} =
\begin{pmatrix}
0\cr
\mathbf{0}\cr
0 
\cr
  w_\m
\cr
 w_\r
 \cr
0\cr
\vdots\cr
0
\end{pmatrix},
\end{equation}

Here we have introduced ``work-like'' quantities
$w_m$ and $w_r$ that represent any changes in the thermal
and radiation energies that are not already included in
the explicit terms of $f_{hyperbolic} + f_{fixup1}$.
In numerical application, we first apply the
updates $f_{hyperbolic}+f_{fixup1}$  terms to
a discretized version of $\mathitbf{U}^n$ at a time $t_n$
to compute an intermediate state:
$$\Utilde^{n+1}  =  \mathitbf{U}^n +(f_{hyperbolic}+f_{fixup1})\,\Delta t .$$
This is done by first using a (slightly modified) traditional Godunov method
for a conservative update as per $f_{hyperbolic}$, and then applying additional
terms. An important modification is the $- \lambda \nabla \Erad $ term in
the momentum equation. We currently use precomputed $\lambda^{n}$ values
based on the previous time step in the implementation,
represented on the same discrete grid used for cell-centered conservative 
variables. We have implemented numerical spatial smoothing of this flux-limiter variable
to counteract instabilities that we found in some simulations.

For the components of $\mathitbf{U}^n$ we have
$ \Etot = \rho \emat + \Erad +   \rho\frac{\vv^2}{2}$,
this will in general not be true for the components of $\Utilde^{n+1}$,
and we compute the energy mismatch
$$\Delta \Etot = \widetilde{\Etot} - \left(\tilde\rho \tilde{\emat} + \tilde{\Erad} +   \tilde\rho\frac{\vvtilde^2}{2}\right),$$
where tilde indicates components of $\Utilde^{n+1}$.

Next we reestablish consistency between the energy components by applying the $f_{fixup2}$ term.
Note that we trust the value of $\widetilde{\Etot}$ (as well as $\tilde\rho$ and $\vvtilde^2$), 
which come from the conservative update of the hyperbolic system, more than the updated values of
$\tilde{\emat}$ and $\tilde{\Erad}$, so we adjust the latter by partitioning the
energy mismatch among then, such that $\Delta \Etot = \left( w_\m + w_\r  \right) \Delta t$.
We have implemented various strategies for effecting this partitioning. 
We briefly describe here ``RAGE-like energy partitioning'' (RLEP), which is based on the
same approach that has been implemented in the FLASH code (Release 4 and later) for partitioning of energies
between electron and ion components, which in turn is described in \cite{2008CS&D....1a5005G}.

Let $q_c = w_c \Delta t$ for $c\in\{{\mathrm m},{\mathrm r}\}$.  Let
$\pmat^+$, $\Pradeff^+ = \lambda \tilde{\Erad}$, and
$\ptot^+ = \pmat^+ + \Pradeff^+$
 be predicted values of matter, effective radiation, and total pressures, respectively,
at time $t_{n+1}$ that can be computed by \code{Eos} calls on the $\Utilde^{n+1}$ state.
Then define
\begin{eqnarray}
q_\m &=& \frac{\pmat^+}{\ptot^+}  \Delta \Etot,\quad   q_\r = \frac{\Pradeff^+}{\ptot^+}  \Delta \Etot,
\end{eqnarray}
\ie, simply partition the energy mismatch in proportion to the pressure ratios.
We also use additional fallbacks and heuristics, \eg, to recover from unphysical nonpositive energy values.

\end{document}